\documentclass[journal]{IEEEtran}
\usepackage{amsmath, amssymb, amsfonts}
\usepackage{algorithmic}
\usepackage{graphicx}
\usepackage{array}
\usepackage{url}
\usepackage{cite}
\usepackage{multirow}
\usepackage{color}
\usepackage{bm}
\usepackage{subfigure}
\usepackage{siunitx}
\IEEEoverridecommandlockouts

\begin{document}
\title{An Overview on IEEE 802.11bf: WLAN Sensing}
	
\author{
	Rui Du\IEEEauthorrefmark{1}, \IEEEmembership{Member, IEEE,}
	Hailiang Xie\IEEEauthorrefmark{1}, \IEEEmembership{Graduate Student Member, IEEE,}
	Mengshi Hu, 
	Narengerile,
	Yan Xin,
	Stephen McCann, \IEEEmembership{Senior Member, IEEE,}
	Michael Montemurro,
	Tony Xiao Han, \IEEEmembership{Senior Member, IEEE,}
	and Jie Xu, \IEEEmembership{Senior Member, IEEE}
	\thanks{
		Rui Du, Mengshi Hu, Narengerile, Yan Xin, Stephen McCann, Michael Montemurro, and Tony Xiao Han are with Huawei Techologies Co., Ltd. (email: tony.hanxiao@huawei.com).}
	\thanks{
		Hailiang Xie is with the Future Network of Intelligence Institute (FNii), The Chinese University of Hong Kong (Shenzhen), Shenzhen, China, and the School of Information Engineering, Guangdong University of Technology, Guangzhou, China (e-mail: hailiang.gdut@gmail.com).}
	\thanks{
		Jie Xu is with the School of Science and Engineering (SSE), the Future Network of Intelligent Institute (FNii), and the Guangdong Provincial Key Laboratory of Future Networks of Intelligence, The Chinese University of Hong Kong (Shenzhen), Shenzhen 518172, China (e-mail: xujie@cuhk.edu.cn). Tony Xiao Han and Jie. Xu are the corresponding authors. }
	\thanks{
		\IEEEauthorrefmark{1}Co-first authors.}
	}

\maketitle

\begin{abstract}
With recent advancements, the wireless local area network (WLAN) or wireless fidelity (Wi-Fi) technology has been successfully utilized to realize sensing functionalities such as detection, localization, and recognition. 
However, the WLANs standards are developed mainly for the purpose of communication, and thus may not be able to meet the stringent sensing requirements in emerging applications. 
To resolve this issue, a new Task Group (TG), namely IEEE 802.11bf, has been established by the IEEE 802.11 working group, with the objective of creating a new amendment to the WLAN standard to provide advanced sensing requirements while minimizing the effect on communications. 
This paper provides a comprehensive overview on the up-to-date efforts in the IEEE 802.11bf TG. First, we introduce the definition of the 802.11bf amendment and its standardization timeline. 
Then, we discuss the WLAN sensing procedure and framework used for measurement acquisition, by considering both conventional sensing at sub-7 GHz and directional multi-gigabit (DMG) sensing at 60 GHz, respectively. 
Next, we present various candidate technical features for IEEE 802.11bf, including waveform/sequence design, feedback types, quantization, as well as security and privacy. 
Finally, we describe the methodologies used by the IEEE 802.11bf TG to evaluate the alternative performance. 
It is desired that this overview paper provide useful insights on IEEE 802.11 WLAN sensing to people with great interests and promote the IEEE 802.11bf standard to be widely deployed.	
\end{abstract}

\begin{IEEEkeywords}
	IEEE 802.11bf, WLAN sensing, Wi-Fi sensing, DMG sensing.
\end{IEEEkeywords}

\section{Introduction}
Over the recent 20 years, wireless fidelity (Wi-Fi$^\circledR$) has evolved from a nascent wireless local area network (WLAN) technology based on the family of IEEE 802.11 standards to a necessity in business and life around the world. 
As reported by the Wi-Fi Alliance$^\circledR$ \cite{WiFi_Alliance}, the global economic value provided by Wi-Fi reached $\$3.3$ trillion in 2021 and is expected to grow to nearly $\$5$ trillion by 2025. 
Besides the social and economic benefits, Wi-Fi has also led to the continuous innovation and development to enable a wide range of new services for supporting emerging applications. 
Among others, \textit{WLAN sensing}, also known as {\it Wi-Fi sensing}\footnote{Wi-Fi has gradually become synonymous with WLAN due to its simplicity, reliability, and flexibility. In the 802.11 standards group, ``Wi-Fi sensing'' can be considered as an equivalent term of ``WLAN sensing''. Therefore, this paper uses ``Wi-Fi sensing'' and ``WLAN sensing'' interchangeably.}, has recently attracted growing interests from both academia and industry. 

\begin{figure}[htbp]
	\centering
	\includegraphics[width=1\linewidth]{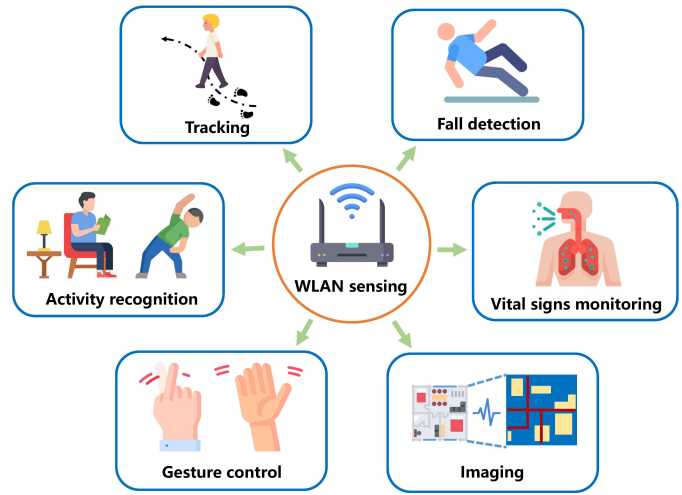}
	\caption{Example applications of WLAN sensing.}
	\label{fig_Sensing}
\end{figure}

WLAN sensing is a technology that uses Wi-Fi signals to perform sensing tasks, by exploiting prevalent Wi-Fi infrastructures and ubiquitous Wi-Fi signals over surrounding environments. 
In particular, Wi-Fi radio waves can bounce, penetrate, and bend on the surface of objects during their propagation. 
By proper signal processing, the received Wi-Fi signals can be harnessed to sense surrounding environments, detect obstructions, and interpret target movement. 
By implementing this, WLAN sensing has been successfully used in abundant residential, enterprise, indoor, and outdoor applications as shown in Fig. \ref{fig_Sensing}, including gesture control, fall detection, tracking, imaging, activity recognition, vital signs monitoring, etc. 
For instance, the Cognitive Systems Corp. released a commercial product using WLAN sensing for motion detection \cite{cognitive}, and Origin Wireless, a start-up company, focused on the commercialization of WLAN sensing with the goal of bringing WLAN sensing to the world by partnering with key players in major verticals \cite{originwireless}. 
Therefore, WLAN sensing creates many new opportunities for Wi-Fi service providers to enter these markets.


The research on WLAN sensing can be traced back to the beginning of the 21st century, and its feasibility has been independently demonstrated by different groups in both academia and industry\cite{RSSI1,RSSItoCSI,survey_ILS,RSSI_fingerprint,RSSI_PL_model,CSI_survey, SENS_tutorial}. 
In general, the WLAN sensing can be classified into two main categories, which are implemented based on different wireless signal characteristics, namely the received signal strength indicator (RSSI) and channel state information (CSI). 
Specifically, the RSSI corresponds to the measured received signal strength at the receiver\footnote{In fact, RSSI is a relative measure of the actual received signal strength (RSS), for which the relative magnitude can be freely defined by different chip suppliers \cite{survey_ILS}. For ease of exposition, in this paper we uniformly use RSSI to denote the actual measured received signal strength.} that has been widely used in the early attempts of WLAN sensing \cite{RSSI1,RSSItoCSI} based on fingerprint and geometric model based methods. 
For example, RSSI patterns at different locations can be used as fingerprints for localization \cite{RSSI_fingerprint}. By employing a simple path loss model, RSSI can also be used to estimate the distance between the transmitter and receiver \cite{RSSI_PL_model}. 
In general, the RSSI-based approaches are easy to implement and with low cost. 
However, as the RSSIs may fluctuate significantly over time and space due to the multi-path effect in complex environments and the imperfect automatic gain control (AGC) circuit at Wi-Fi devices, such approaches may result in degraded sensing performance. 
Different from RSSI, CSI is able to provide finer-grained wireless channel information at the physical layer, which is thus considered as an alternative solution for accurate sensing\cite{CSI_survey}. 
CSI contains both channel amplitude and phase information over different subcarriers that provide the capability to discriminate multi-path characteristics. 
For instance, by processing the spatial-, frequency-, and time-domain CSI at multiple antennas, subcarriers, and time samples via fast Fourier transform (FFT), we can extract detailed multi-path parameters such as angle-of-arrival (AoA), time-of-flight (ToF), and doppler frequency shift (DFS). 
Other advanced super-resolution techniques such as estimation of signal parameters via rotation invariance techniques (ESPRIT)\cite{ESPRIT}, multiple signal classification (MUSIC)\cite{MUSIC}, and space alternating generalized expectation-maximization (SAGE) algorithm\cite{SAGE} can also be utilized to extract  more accurate target-related parameters from the CSI. 
In various prior works \cite{WiSH, Chronos,IndoTrack,Widar2,Wi-Sleep,TVS,PhaseBeat,MultiSense}, the CSI-based sensing approaches have been demonstrated to provide high sensing accuracy for detection and tracking.

\begin{table*}[htbp]
	\caption{Wireless Technologies Overview}
	\centering
	\begin{tabular}{c c c c c c c}
		
		\hline\hline & & & & & & \\ [-1.5ex]
		Technology	& Coverage & Power consumption & Dedicated devices & Cost & Accuracy & Disadvantages \\ [0.5ex]
		\hline & & & & & & \\ [-1.5ex]
		Visible Light	& Room 	& Low		& Yes & Moderate		& Low-moderate	& Need LOS scenario	\\
		Ultrasound		& Room    & High		& Yes & Moderate-high	& Moderate-high	& Interference	\\
		RFID			& Room 	& Low		& Yes & Low				& Low-moderate	& Response time is high	\\
		UWB				& Building & Moderate	& Yes & High			& High			& Lack of well-developed infrastructure	\\
		Bluetooth		& Building & Low		& Yes & Low-moderate	& Low-moderate	& Limited coverage	\\
		Wi-Fi			& Building	& Moderate	& No  & Low				& Moderate-high & Need standard modification \\ [0.5ex]	
		\hline
	\end{tabular}
	\label{table_tech}
\end{table*}

WLAN sensing differs significantly from other existing sensing technologies based on, e.g., visible light, ultrasound, radio frequency identification (RFID), Bluetooth, and ultra-wideband (UWB). 
In particular, visible light positioning (VLP) estimates the locations of light sensors based on the visible lights transmitted from the light-emitting diode (LED) transmitters at known locations, which highly depends on the line-of-sight (LOS) channel between transmitters and receivers\cite{VLP1}. 
By contrast, WLAN sensing can provide better coverage as radio signals can propagate through walls to provide additional non-LOS (NLOS) information. 
Furthermore, VLP needs dedicated infrastructures such as photodetectors and imaging sensors that may lead to high system cost \cite{VLP2,VLP3}, while WLAN sensing can reuse existing Wi-Fi devices with significantly less cost. 
Next, ultrasound-based sensing uses an ultrasonic transceiver to record the ToF between the transmitter and receiver, and then calculates their separation distance based on the speed of sound \cite{UPS1}. 
However, unlike Wi-Fi signals that are not harmful to health, the ultrasonic signals can be harmful to infants and pets that are sensitive to high-frequency sounds \cite{UPS2}. 
In addition, the speed of sound varies significantly with humidity and temperature (e.g., the speed of sound may increase by about 0.6 m/s for every one degree increase in temperature), thus making it less stable than WLAN sensing based on electromagnetic wave. 
Furthermore, RFID requires a separate infrastructure to be deployed \cite{RFID1}, whereas Wi-Fi is able to leverage network access infrastructure that is already deployed, which thus allows for large-scale commercial use. 
Moreover, RFID systems need to deploy dedicated arrays of passive RFID tags at targets, which further increases the deployment cost \cite{RFID2}. 
In addition, Bluetooth is a wireless personal area network (WPAN) technology to enable short-range wireless communication. 
As compared to Wi-Fi, Bluetooth has lower throughput, shorter transmission range, and needs to be deployed as a separate infrastructure, thus making the large-scale deployment of Bluetooth-based sensing difficult if not impossible. 
Finally, UWB positioning technologies transmit extremely short pulses over a large bandwidth ($>$500MHz) to track objects in a passive manner \cite{UWB}, which are therefore robust against multi-path issues for obtaining an accurate ToF estimation. However, the lack of large-scale infrastructure deployment limits the utilization of UWB for sensing. In summary, the comparison between the above wireless sensing technologies and WLAN sensing is listed in Table \ref{table_tech}.

\begin{figure*}[htbp]
	\centering
	\includegraphics[width=1\linewidth]{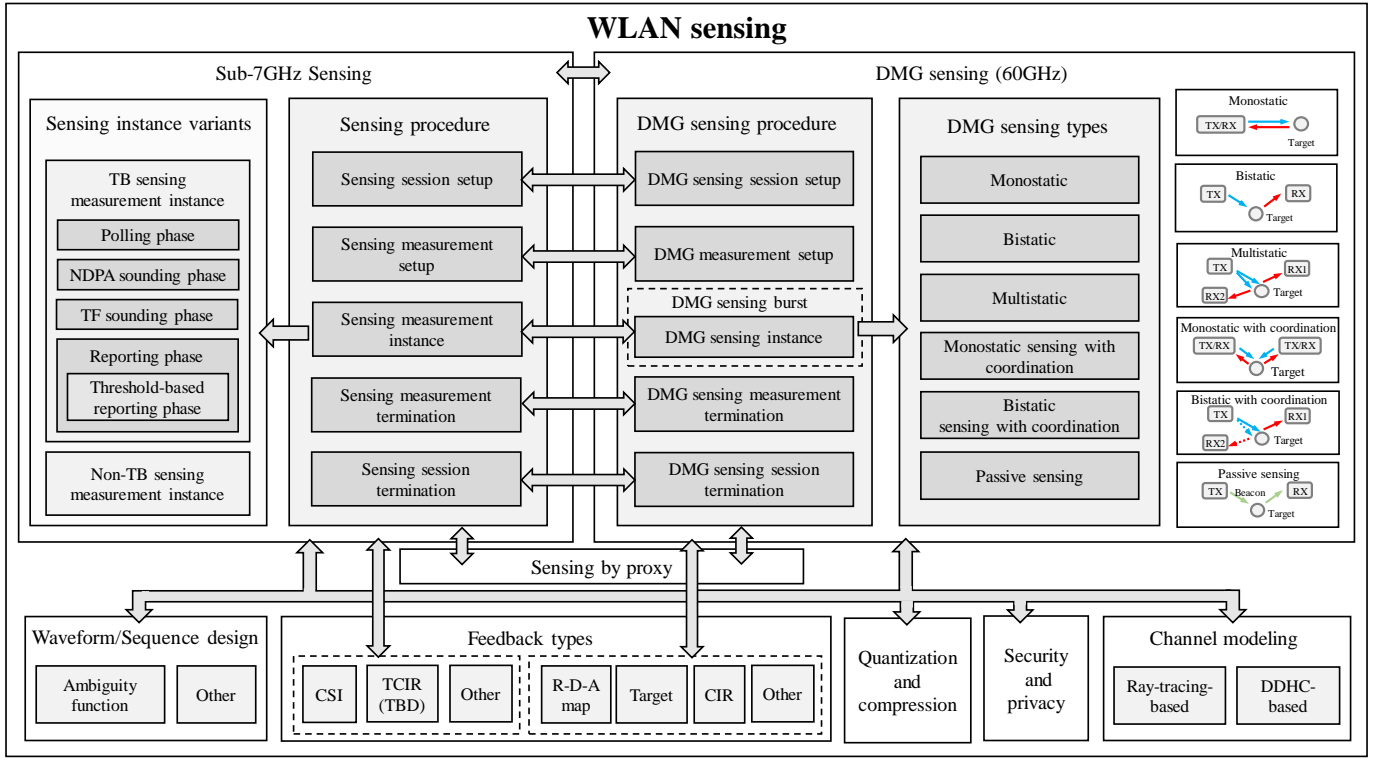}
	\caption{Overview of the WLAN sensing system.}
	\label{WLAN_sensing_system}
\end{figure*}

Despite the benefits and recent advancements, existing WLAN sensing works based on commercial off-the-shelf (COTS) devices still face the following key challenges. This thus motivates the development of new WLAN standards to better support WLAN sensing. 
\begin{itemize}
\item \textit{CSI Availability:} Although IEEE 802.11-compliant CSI extraction methods already exist, CSI is not always available from the user side in commercial COTS Wi-Fi devices. This is because most Wi-Fi chipset manufacturers keep CSI access as a private feature and do not provide a dedicated function interface to consumers. Only a few commercial Wi-Fi devices using the outdated communication standard like 802.11n are able to access CSI data but with very limited flexibility, such as the Intel 5300\cite{Intel5300} and Atheros AR9580\cite{AR9589} Wi-Fi cards. This has undoubtedly hindered the research and development of WLAN sensing. As a result, there is a need to identify a unified and flexible functional interface for channel measurements.

\item \textit{Transmission adaptation:} In WLAN systems, the transceivers may dynamically change the transmission and/or receiving strategies (e.g., the antenna configuration, transmit power, and the AGC levels) based on the channel conditions. If the transmitter and/or receiver does not know {\it a priori} the system parameters such as the number of spatial streams, the number of antennas, and the transmit power, then the sensing measurements may become unreliable, resulting in significant performance degradation. Therefore, proper protocols to support the negotiation between transceivers are required.

\item \textit{Only single-node sensing supported:} In certain scenarios, there may exist a single (or very few) access point (AP) and multiple non-AP stations (STAs), the latter of which can assist in sensing of the former. In general, their sensing collaboration can provide additional information (e.g., more multipaths as well as coverage) and therefore greatly improve the sensing performance. In addition, Wi-Fi devices from distinct vendors may have different sensing and computing capabilities as well as access interfaces. Through unified standardization, cooperation and interaction among multiple STAs or between APs and STAs ( or even between devices from different vendors) can be exploited to enable networked sensing and computing.

\item \textit{Effects on communication:} Traditional WLAN networks are originally designed for data transmission, without considering any sensing function. In this case, the link throughput and latency are widely adopted performance metrics, which, however, are no longer applicable to measure the performance of WLAN sensing. In addition, received wireless signals may suffer from distortions in both amplitude and phase, and such distortions can be compensated in part in conventional communication standards in the equalization phase \cite{phase_error}. However, the accuracy of this phase error compensation is coarse and may not meet the sensing requirements. Thus, enhanced transmission protocol and sounding processes are necessary to improve the WLAN sensing performance, and also enable efficient and reliable integrated sensing and communication (ISAC) \cite{ISAC,ISAC2,ISAC4}.
\end{itemize}

To solve the aforementioned issues above, it is emerging to have new WLAN sensing standards to support sensing functions on Wi-Fi devices in a timely and efficient manner without (significantly) affecting communication performance\footnote{Notice that previous WLAN standards from IEEE 802.11a\textsuperscript{TM}-2009 to the IEEE 802.11be\textsuperscript{TM} for sub-7 GHz to IEEE 802.11ay-2021 for 60 GHz mainly focused on communication performance enhancement.}. 
Towards this end, IEEE 802.11 is expected to release the WLAN sensing standard amendment, i.e., IEEE 802.11bf, for WLAN sensing and ISAC. 
This amendment defines standardized modifications to the IEEE 802.11 physical (PHY) layer and medium access control (MAC) layer that not only enhance the sensing capability but also lead to the ease of deployment. 
It is worth noting that the IEEE 802.11bf amendment makes only MAC layer modifications for the sub-7GHz band, but both PHY layer and MAC layer modifications for the 60GHz band. 
Specifically, IEEE 802.11bf supports the 60GHz band sensing by improving and modifying the directional multi-gigabit (DMG) implementation in IEEE 802.11ad-2012 and the enhanced DMG (EDMG) implementation in IEEE 802.11ay-2021, both of which use beamforming to provide higher data rates.
By defining specific standards support, the reliability and efficiency of WLAN sensing can be improved, stimulating further innovation and enabling more applications. 
Furthermore, the development of the standard typically results in the participation of experts from industry stakeholders. 
Although members from different companies have different points of interest, the sharing of information and experience can help standardize and advance the technology. 

It is worth noting that IEEE Standardization Association previously established standards on wireless ranging, namely IEEE 802.11az \cite{11az_PAR}, which was intended to address the need for a station to identify its absolute and relative positions to another station. 
However, IEEE 802.11az requires the object or measurement target to carry a hardware device (i.e., equipped with a cooperating device) for wireless ranging. 
In contrast to this, WLAN sensing focuses on device-free sensing, without requiring any devices or tags to be attached to the targets. 
Additionally, while IEEE 802.11az only focuses on indoor localization and positioning, WLAN sensing considers a much wider range of sensing applications beyond indoor. 

In viewing of the fast development, this paper aims to provide a comprehensive and up-to-date overview on the IEEE 802.11bf standard, by introducing the latest standardization progress. 
In summary, Fig. \ref{WLAN_sensing_system} provides an overview of the current IEEE 802.11bf standardization efforts, which are also the main focus of this paper.

The remainder of this paper is organized as follows. 
Section II introduces how IEEE 802.11bf was formed and the timeline for its standardization. 
Section III presents key use cases for IEEE 802.11bf. 
Section IV describes the sensing procedure in IEEE 802.11bf to address the measurement acquisition problem in sensing. 
Section V discusses the candidate techniques for IEEE 802.11bf. 
Section VI introduces the evaluation methodology and channel model for the IEEE 802.11bf task group. 
Finally, Section VII concludes this paper.

For better understanding, the main terms and acronyms used in this paper are listed alphabetically in Table \ref{Table:Acronyms}. 

\begin{table}[]
	\renewcommand\arraystretch{1.07}
	\centering
	\caption{Summary of Main Acronyms.}
	\begin{tabular}{|c|m{180pt}|}
		\hline
		AAF    & Auto Ambiguity Function                                            \\ \hline
		A-BFT  & Association Beamforming Training                                   \\ \hline
		AF     & Ambiguity Function                                                 \\ \hline
		AGC    & Automatic Gain Control                                             \\ \hline
		AID    & Association Identifier                                             \\ \hline
		AoA    & Angle-of-Arrival                                                   \\ \hline
		AP     & Access Point                                                       \\ \hline
		AWV    & Antenna Weight Vector                                              \\ \hline
		BFT    & Beamforming Training                                               \\ \hline
		BI     & Beacon Interval                                                    \\ \hline
		BRP    & Beam Refinement Phase                                              \\ \hline
		BTI    & Beacon Transmission Interval                                       \\ \hline
		CAF    & Cross Ambiguity Function                                           \\ \hline
		CFR    & Channel Frequency Response                                         \\ \hline
		CIR    & Channel Impulse Response                                           \\ \hline
		COTS   & Commercial Off-The-Shelf                                           \\ \hline
		CSD    & Criteria for Standards Development                                 \\ \hline
		CSI    & Channel State Information                                          \\ \hline
		DDCH   & Data-Driven Hybrid Channel Model                                   \\ \hline
		DFS    & Doppler Frequency Shift                                            \\ \hline
		DMG    & Directional Multi-Gigabit                                          \\ \hline
		EDMG   & Enhanced Directional Multi-Gigabit                                 \\ \hline
		ESPRIT & Estimation of Signal Parameters via Rotation Invariance Techniques \\ \hline
		FOV    & Field of View                                                      \\ \hline
		IFFT   & Inverse Fast Fourier Transform                                     \\ \hline
		ISAC   & Integrated Sensing and Communication                               \\ \hline
		I-TXSS & Initiator Transmit Sector Sweep                                    \\ \hline
		KPI    & Key Performance Indicator                                          \\ \hline
		LOS    & Line-of-Sight                                                      \\ \hline
		MAC    & Medium Access Control                                              \\ \hline
		MUSIC  & MUltiple SIgnal Classification                                     \\ \hline
		NDP    & Null Data Packet                                                   \\ \hline
		NDPA   & Null Data Packet Announcement                                      \\ \hline
		NLOS   & Non-Line-of-Sight                                                  \\ \hline
		PAPR   & Peak to Average Power Ratio                                        \\ \hline
		PAR    & Project Authorization Request                                      \\ \hline
		PDP    & Power Delay Profile                                                \\ \hline
		PHY    & Physical                                                           \\ \hline
		PPDU   & Physical Layer Protocol Data Unit                                  \\ \hline
		RCS    & Radar Cross Section                                                \\ \hline
		RFID   & Radio Frequency Identification                                     \\ \hline
		RMSE   & Root Mean Square Error                                             \\ \hline
		RSSI   & Received Signal Strength Indicator                                 \\ \hline
		R-TXSS & Responder Transmit Sector Sweep                                    \\ \hline
		SAGE   & Space Alternating Generalized Expectation-Maximization             \\ \hline
		SBP    & Sensing by Proxy                                                   \\ \hline
		SIFS   & Short Interframe Space                                             \\ \hline
		SISO   & Single-Input Single-Output                                         \\ \hline
		SLAM   & Simultaneous Localization and Mapping                               \\ \hline
		SNR    & Signal-to-Noise Ratio                                              \\ \hline
		SSW    & Sector Sweep                                                       \\ \hline
		STA    & Station                                                            \\ \hline
		Sync   & Synchronization                                                    \\ \hline
		TB     & Trigger-based                                                      \\ \hline
		TCIR   & Truncated Channel Impulse Response                                 \\ \hline
		TF     & Trigger Frame                                                      \\ \hline
		ToF    & Time-of-Flight                                                     \\ \hline
		TPDP   & Truncated Power Delay Profile                                      \\ \hline
		TRN    & Training                                                           \\ \hline
		TRRS   & Time-Reversal Resonation Strength                                  \\ \hline
		UID    & Un-association Identifier                                          \\ \hline
		UWB    & Ultra-Wideband                                                     \\ \hline
		Wi-Fi  & Wireless Fidelity                                                  \\ \hline
		WLAN   & Wireless Local Area Network                                        \\ \hline
	\end{tabular}
	\label{Table:Acronyms}
\end{table}

\section{IEEE 802.11bf: Formation, Definition, and Timeline}
\subsection{Formation}
The formation of a WLAN sensing project was first discussed in the IEEE 802.11 Wireless LAN Next-Generation Standing Committee (WNG SC) in July 2019\cite{WNG_SEN1,WNG_SEN2,WNG_SEN3}, where the feasibilities for WLAN to support sensing use cases and their requirements were justified. 
In October 2019, the WLAN sensing Topic Interest Group (TIG) was initially established. 
In November 2019, the formation of the WLAN sensing Study Group (SG) was approved by the New Standard Committee (NesCom) in the IEEE Standards Association (SA)\cite{SENS_SG}. 
In the WLAN sensing SG, the Project Authorization Request (PAR) and Criteria for Standards Development (CSD) were developed. 
Specifically, a PAR is a formal document that defines the motivation, scope, and content for a proposed standard or amendment. 
Moreover, it is the means by which standards projects are started within the IEEE SA. 
The CSD document is an agreement between the WG and the sponsor, providing a more detailed description of the project and the requirements of the sponsor that is required by the PAR. 
After collaboration, consensus was reached within the WLAN sensing TIG/SG on the PAR and CSD. They were placed on the agenda of the New Standards Committee (NesCom), pending approval by the IEEE Standards Committee. 
With their approval of the IEEE 802.11bf PAR and CSD in September 2020, a new task group (IEEE 802.11bf) about WLAN sensing within the scope of the IEEE 802.11 working group (WG), was officially established, based on which concrete discussions towards creating an amendment for WLAN sensing begin. 

\subsection{Definition}
According to the formal definition of IEEE 802.11bf, WLAN sensing refers to the use of wireless signals received from WLAN sensing capable STAs to determine the features (e.g., range, velocity, angular, motion, presence or proximity, gesture) of the intended targets (e.g., object, human, animal) in a given environment (e.g., room, house, vehicles, enterprise). 

As specified in the PAR, IEEE 802.11bf aims to develop an amendment that defines modifications to the IEEE 802.11 MAC, the DMG and EDMG PHYs to enhance WLAN sensing operation in license-exempt frequency bands between 1 GHz and 7.125 GHz and above 45 GHz. The amendment is expected to enable backward compatibility and coexistence with existing or legacy IEEE 802.11 devices operating in the same band, by providing some basic levels of support for WLAN sensing. The IEEE 802.11bf amendment is expected to have the following features \cite{11bf_PAR}:
\begin{enumerate}
	\item Stations (STAs)\footnote{STA is a device that contains IEEE 802.11 MAC and PHY interfaces to the wireless medium (WM) \cite{802.11_specif}. For example, an STA can be an AP, a laptop, and a WLAN-enabled phone. Therefore, STAs can be mainly divided into AP STAs and non-AP STAs.} to perform one or more of the following: To inform other STAs of their WLAN sensing capabilities, to request and set up transmissions that allow for WLAN sensing measurements, to indicate that a transmission can be used for WLAN sensing, and to exchange WLAN sensing feedback and information;
	\item WLAN sensing measurements to be obtained using transmissions that are requested, unsolicited, or both;
	\item A MAC service interface for layers above the MAC to request and retrieve WLAN sensing measurements.
\end{enumerate}

\begin{figure*}[htbp]
	\centering
	\includegraphics[width=1\linewidth]{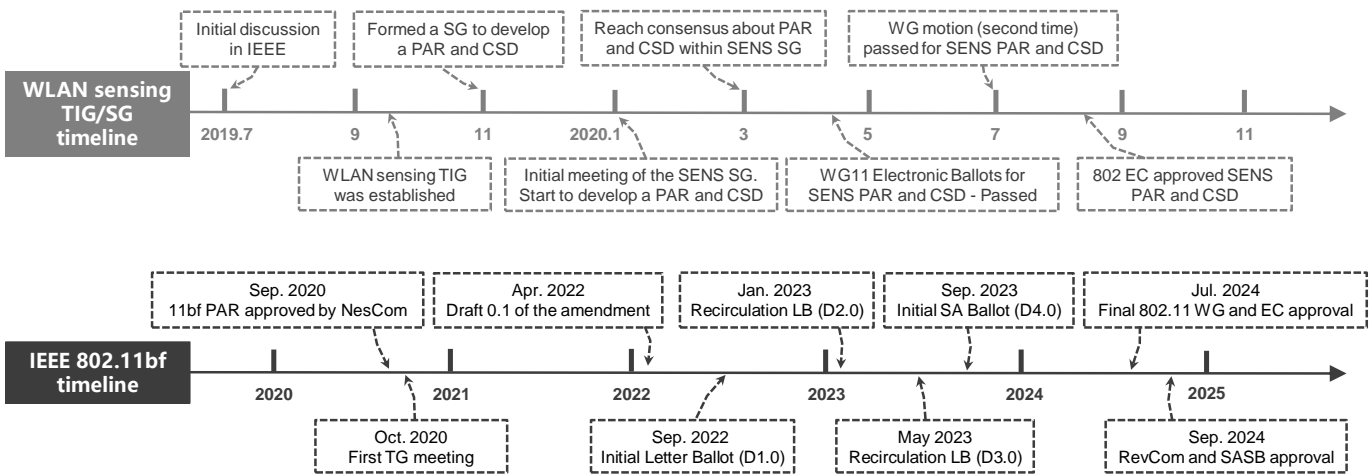}
	\caption{Timeline and progress for the IEEE 802.11bf.}
	\label{11bf_timeline}
\end{figure*}

\subsection{Timeline}
Unlike previous amendments that focused on improving communication metrics such as throughput and latency, the 802.11bf amendment focuses on improving WLAN sensing performance while maintaining or improving a certain level of communication performance. 
Fig. \ref{11bf_timeline} illustrates the timeline and progress towards the completed IEEE 802.11bf amendment. During the TIG/SG phase of this project (September 2019-September 2020), different topics (such as channel model, sequence design, and channel measurement procedure) were presented, and the proposed enhancements have been shown to be technically feasible. 
The first draft of the IEEE 802.11bf amendment (i.e., Draft 0.1) was released by IEEE 802.11bf in April 2022\cite{SENS_SG}. 
Multiple letter ballots (i.e., technical ballots determining whether the draft of the amendment should be approved) have been or will be conducted afterward. 
In each ballot, IEEE 802.11 voting members can vote for or against the draft, and can optionally attach comments for revision. Submission of the draft amendment to the IEEE SA for a subsequent SA ballot is expected as early as September 2023. 
Finally, after the recommendation of the Standards Review Committee (RevCom) and the approval of the IEEE SA Standards Board, the project is expected to be published as an IEEE 802.11bf amendment specification in September 2024. 
Actual deployment of the standard is expected to take place as early as the end of 2024.

\section{Use Cases and Performance requirements}
This section presents the WLAN sensing use cases and identifies the key performance indicator (KPI) requirements. In contrast to the IEEE 802.11 standards, IEEE 802.11bf requires a new variety of KPIs. In particular, a series of KPIs for WLAN sensing have been defined by IEEE 802.11bf, as described below\cite{11bf_KPI}.
\begin{itemize}
	\item \textbf{Range Coverage:} The maximum allowable distance from a sensing STA to the target, within which the signal-to-noise ratio (SNR) is above a pre-defined threshold (conventionally taken as 10dB or 13dB), such that the targets can be successfully detected.
	\item \textbf{Field of View (FOV):} The angle through which the STA performs sensing, i.e., the FOV indicates the coverage area of a sensing device in terms of angle.
	\item \textbf{Range Resolution:} The minimum distance between two targets that a sensing STA can distinguish on the same direction but at different ranges. 
	\item \textbf{Angular Resolution (Azimuth / Elevation):} The minimum angle between two targets at the same range, such that the sensing STA is able to distinguish.
	\item \textbf{Velocity Resolution:} The minimal velocity difference between two objects that a sensing STA can distinguish.
	\item \textbf{Accuracy:} The difference between the estimated range/angle/velocity of an object and the ground truth.
	\item \textbf{Probability of Detection:} The ratio of the number of correct predictions to the number of all possible predictions. The prediction tasks can be:
	\begin{itemize}
		\item[a)] gesture detection, where a pre-defined set of gestures and/or motions shall be identified;
		\item[b)] presence detection; 
		\item[c)] a specific body activity detection like breathing; 
		\item[d)] real person detection, distinguishing human beings from other objects.
	\end{itemize}
	\item \textbf{Latency:} Expected time taken to finish the related WLAN sensing process.
	\item \textbf{Refresh Rate:} Frequency when the sensing refresh takes place.
	\item \textbf{Number of Simultaneous Targets:} The number of targets that can be detected simultaneously within the sensing area.
\end{itemize}

Based on the main application scenarios for WLAN sensing, IEEE 802.11bf defines a variety of use cases in \cite{11bf_use_case}, where the performance requirements for each use case are also outlined. For different use cases, various WLAN sensing designs have been studied to improve sensing performance. Here, we briefly discuss a few key use cases.

\subsubsection{Presence Detection}
In a typical indoor scenario, reliable human presence detection is key to achieving smart home (e.g., home control). 
This is significant for preventing energy waste and improving the user experience. 
The presence detection can be mainly divided into two different states: moving (e.g., walking or making large movements) and stationary (e.g., lying, sitting, or standing still). 
As the variation of CSI in the time domain has different patterns for humans in different states, these patterns can be used for presence detection. 
In addition, some other features can be exploited to enhance the differentiation between different states, such as time-reversal resonation strength (TRRS)\cite{TRRS}, correlation \cite{WiSH}, higher-order moments, and Doppler spectrum. 
The key challenge of presence detection lies on the detection of a stationary human, which is due to the fact that the measured CSI with a stationary human is generally similar to that in empty rooms dominated by white Gaussian noise. 
According to \cite{11bf_use_case}, the maximum range coverage requirement for presence detection is 10-15 m, and specific value needs to be selected according to the room size. 
For example, in order to detect the presence and count the number of people in a meeting room, the maximum range coverage requirement can be generally set as 10 m. 
In addition, for presence detection in a multi-person environment, the range resolution, velocity resolution, and angular resolution need to be at least 0.5-2 m, 0.5 m/s, and 4-6 degrees, respectively.

\subsubsection{Activities Recognition}
Human activity recognition (HAR) plays a significant role in human-computer interaction (HCI) to help the computer understand human behaviors and intention. 
HAR with Wi-Fi has been used in various applications, such as fall detection, gesture recognition, and security. 
As a wireless channel can be distorted by human activities, we can extract patterns such as the Doppler spectrum, target speed, and amplitudes from the estimated CSI, in order to detect or to recognize human daily activities. 
However, there are two challenges to tackle. 
First, the same activity may generate different patterns at different places, since each receiver can only record radial velocity of targets. 
Next, most existing prototype systems assumed the existence of an obvious pause between adjacent activities for segmenting them. 
This, however, is not true, as the human activities are usually continuous without pause. 
This thus makes it challenging to automatically segment continuous activities \cite{wifi isac challenges}. 
IEEE 802.11bf allows the use of the sub-7 GHz band for large motion (i.e., full-body motion) recognition and the 60 GHz band for small motion (i.e., finger and hand motion) recognition to provide higher resolution and improved recognition accuracy \cite{11bf_use_case}.

\subsubsection{Human Target Localization and Tracking}
Localization is the process of determining the position of a target in the region of interest, while tracking aims to confirm the trajectory of movement using the change in position of the target over time. 
Existing methods for indoor human target tracking/localization can be divided into two main categories: fingerprinting-based and geometric model-based methods. 
Due to its simplicity and deployment practicability, fingerprinting-based localization is one of the most widely used techniques for realizing indoor localization, but the construction of the fingerprint database is troublesome and time-consuming. 
Compared with fingerprinting, the geometric model-based human tracking method has better generalization in different environments. 
In this method, the super-resolution joint multi-parameter estimation algorithm can be adopted to improve multipath resolution to enable human target localization and tracking \cite{Chronos,IndoTrack,Widar2}. 
However, the accuracy of parameter estimation can still be limited due to the small number of antennas and limited bandwidth. 
The KPI for human target localization and tracking is range accuracy within 0.2 m\cite{11bf_use_case}. To support such a high range accuracy, the 60 GHz band sensing is needed.

\subsubsection{Healthcare}
Respiration and heartbeat estimation are two common techniques to analyze human health conditions, such as sleep quality analysis\cite{Wi-Sleep}. 
By observing the vibration pattern of the phase and/or amplitude of the CSI, it is possible to estimate the respiration or heartbeat rate. 
Prior works such as TVS\cite{TVS} and PhaseBeat\cite{PhaseBeat} validated its technical feasibility. 
Nevertheless, estimating the respiration rate for multiple targets is still challenging. 
MultiSense\cite{MultiSense} was recently proposed for estimating multiple persons’ respiration rates, where the multi-person respiration sensing is treated as a blind source separation (BSS) problem. 
The breathing rate accuracy and pulse accuracy are the most important KPIs for respiration and heartbeat estimation. 
In addition, sneeze sensing has emerged as another important sensing use case, which plays a key role in transferring respiratory diseases such as COVID-19 between infectious and susceptible individuals. 
It is konwn that Doppler analysis of a spray droplets cloud is possible at high frequency (60 GHz) by calculating the volume radar cross section (RCS) for the sneeze droplets \cite{Wi-Sneeze}. 
For such applications, at least 0.1 m/s velocity accuracy is generally needed based on the Doppler frequency estimation \cite{11bf_use_case}. 

\subsubsection{3D Vision} 
3D environment reconstruction and 3D perception of human skeletons are two typical 3D vision applications, that transform the surrounding environment and sensing targets into 3D knowledge models, respectively. 
Primitve methods for 3D vision are normally based on LiDARs (e.g., simultaneous localization and mapping (SLAM)), computer vision (CV) (e.g., OpenPose API\cite{openpose} for human sensing), or both. 
Thanks to the sparse channel brought by millimeter wave (mmWave), several initial attempts towards building a 2D map of the surrounding environment based on WLAN sensing have been conducted. For instance, a 2D range angle chart was measured in a corridor using 28 GHz mmWave by solving least square estimation problems\cite{mmwave_mapping}. 
A potential radio SLAM algorithm was proposed in \cite{slam}, which is implemented via the execution of a message passing algorithm over a factor graph, where environment features are contained in collected multipath components. 
As for perception of human skeletons, a 3D skeleton (14 body joints) was sensed initially in \cite{rf_skeleton} using radio signals (5.4-7 GHz frequency modulated continuous wave (FMCW) signals). 
Inspired by this, there are several follow-up items of work using CSI-based WLAN sensing and machine learning techniques to estimate 3D skeletons\cite{person_in_wifi,wipose}. 
As shown in \cite{11bf_use_case}, 3D vision is only applicable to the 60 GHz band, and its KPI is the accuracy of the 3D map. 
Towards this end, the range accuracy, velocity accuracy, and angular accuracy need to be at least 0.01 m, 0.1 m/s, and 2 degrees, respectively, which are the most demanding among all use cases.

Although a great deal of research for each use case is available, there is still a gap between their current performance and the KPIs required. One of the main reasons is that the vast majority of current research on WLAN sensing has been conducted based on existing communication standards to obtain channel measurement information, which may lead to an unstable sensing process and affect the communication functionality. This also leads to a lack of negotiation and cooperation among multiple STAs involved in sensing, making it difficult to further improve the sensing performance of the network. By defining unified and standardized WLAN sensing-specific operations, the 802.11bf standard will maximize the features, efficiencies, and capabilities of WLAN sensing for future improvements in sensing performance. It not only enables sensing STAs to behave in a specific, deterministic way but also enables the network to operate jointly to support a variety of sensing use cases. Furthermore, the advent of 802.11bf provides a platform for the introduction of various innovative sensing algorithms.

\section{IEEE 802.11bf Sensing Framework and Procedure}
From a standardization perspective, the most essential issue to be dealt within IEEE 802.11bf is measurement acquisition, where the goal is to obtain sensing measurements from an IEEE 802.11-based radio. 
Since different use cases require different requirements, 802.11bf defines two different sensing measurement acquisition procedures that can operate in the sub-7 GHz and 60 GHz bands, respectively. 
For ease of exposition, in the sequel, we refer to sub-7 GHz sensing as WLAN sensing, and 60 GHz sensing as DMG sensing. 
As the DMG sensing additionally implements the directional beamforming, more sophisticated design is required.
In this section, we first introduce the basic IEEE 802.11bf definition for a sensing STA and then provide details of the two sensing procedures to explain how measurement acquisition is achieved.

\subsection{Transceiver Roles and Specific Definitions}
To start with, IEEE 802.11bf defines two terminologies, namely a sensing procedure and a sensing session, where the sensing procedure allows a STA to perform WLAN sensing and obtain measurement results, and the sensing session is an agreement between a sensing initiator and a sensing responder to participate in the sensing procedure. 
Here, the sensing initiator and sensing responder are defined depending on which STA initiates a WLAN sensing procedure, and requests and/or obtains measurements. 
A sensing initiator is a STA that initiates a sensing procedure, while a sensing responder is a STA that participates in a sensing procedure initiated by a sensing initiator. Both the sensing initiator and sensing responder can be an AP or a non-AP STA (i.e., client). 

On the other hand, depending on who transmits the IEEE 802.11-based signal (i.e., physical layer protocol data unit (PPDU)\footnote{It is noticed that a PPDU is a data unit exchanged between two peer PHY entities to provide the PHY data service.}) used to obtain measurements, two other types of roles are also available namely sensing transmitter and sensing receiver, respectively. 
Specifically, a sensing transmitter is a STA that transmits PPDUs used for sensing measurements in a sensing procedure and a sensing receiver is a STA that receives PPDUs sent by a sensing transmitter and performs sensing measurements in a sensing procedure. 
The ability to define the role of each STA as sensing transmitter or sensing receiver is an important feature of WLAN sensing.

\begin{figure}[htbp]
	\centering
	\includegraphics[width=1\linewidth]{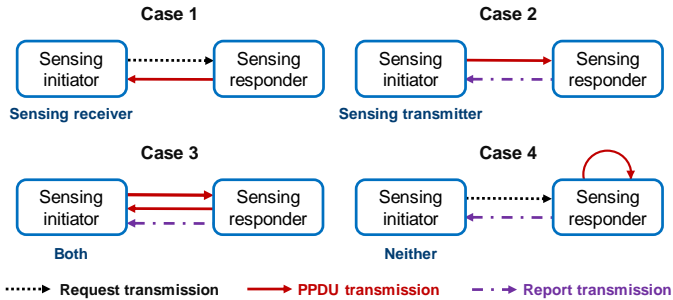}
	\caption{Sensing configuration for IEEE 802.11bf.}
	\label{11bf_SENS_Config}
\end{figure}

It is noticed that a sensing initiator can be either a sensing transmitter or a sensing receiver, both or neither, during a sensing procedure. 
The sensing responder is similar to the sensing initiator and can be either the sensing transmitter, the sensing receiver or both. 
In addition, a STA can assume multiple roles in one sensing procedure. 
Taking the sensing initiator as an example, the sensing configuration used by WLAN sensing can be roughly divided into four cases, as shown in Fig. \ref{11bf_SENS_Config}, which are explained in the following. 
\begin{itemize}
	\item In the first case, the sensing initiator is the sensing receiver, which directly obtains measurements by itself using PPDUs transmitted by the sensing responder.
	\item In the second case, the sensing initiator is the sensing transmitter, which transmits PPDUs and performs the sensing function by using the feedback of measurements from the sensing responder.
	\item In the third case, the sensing initiator is both a sensing transmitter and a sensing receiver, which can obtain uplink measurements by receiving PPDUs and obtain downlink measurements through feedback.
	\item Finally, the sensing initiator is neither a sensing transmitter nor a sensing receiver. In this case, the sensing responder feeds back the measurements obtained by other means to the sensing initiator, thus allowing the sensing initiator to obtain the measurements without sensing PPDUs.
\end{itemize}
    A more detailed packet exchange process will be illustrated later.

\begin{figure}[htbp]
	\centering
	\includegraphics[width=0.9\linewidth]{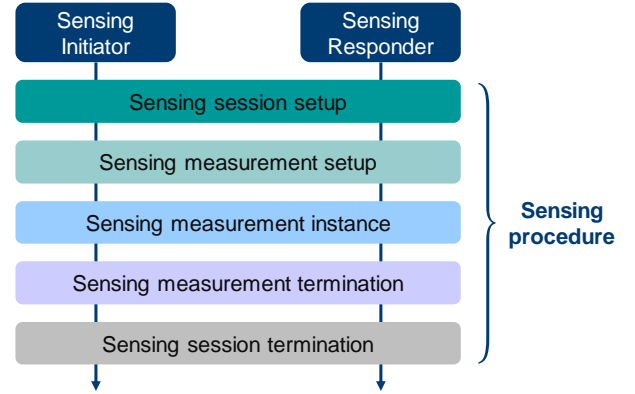}
	\caption{Overview of the WLAN sensing procedure.}
	\label{11bf_SENS_protocol}
\end{figure}

\subsection{General Overview}
A main contribution of the IEEE 802.11bf amendment will be the specification of procedures that allow for WLAN sensing applications to reliably and efficiently obtain measurement results. 
Specifically, through the sensing procedure, IEEE 802.11bf can provide a service that enables a STA to obtain sensing measurements of the channel between two or more STAs and/or the channel between a receive antenna and a transmit antenna of a STA. 
The framework basis of the IEEE 802.11bf WLAN sensing procedure for sub-7 GHz systems is illustrated in Fig. \ref{11bf_SENS_protocol}. 
Specifically, a WLAN sensing procedure typically contains five phases, namely the sensing session setup, sensing measurement setup, sensing measurement instance, sensing measurement termination, and sensing session termination. 
Note that IEEE 802.11bf standardization is ongoing. In the following, details of the general WLAN sensing procedure which have been developed in IEEE 802.11bf so far are introduced.

\subsubsection{Sensing Session Setup}
In the sensing session setup, the sensing initiator establishes a sensing session with the sensing responder(s), and the sensing-related capabilities are exchanged between them at this stage. Multiple sensing sessions can exist simultaneously, where each sensing session needs to be uniquely identified by the MAC address and/or the association identifier (AID)/un-association identifier (UID)\footnote{The AID/UID is a unique number used to identify the transmission between the AP and the associated/un-associated STA.} of the STA establishing the sensing session. In addition, multiple sensing sessions can be maintained by the same sensing initiator to meet the requirements of the WLAN sensing procedure.

\subsubsection{Sensing Measurement Setup} Sensing measurement setup allows for a sensing initiator and a sensing responder to exchange and agree on operational attributes (i.e., special operational information) associated with a sensing measurement instance, which includes the role of the STA, the type of measurement report, and other operational parameters. 
To identify a specific set of operational attributes, measurement setups with different sets of operational attributes are assigned by different Measurement Setup IDs.



\subsubsection{Sensing Measurement Instance}
In the sensing measurement instance, sensing measurements are performed. The Measurement Instance IDs may be used to identify different sensing measurement instances. 

\subsubsection{Sensing Measurement Termination}
In the sensing measurement termination, the corresponding sensing measurement setups are terminated. The sensing initiator and the sensing responder release the allocated resources to store the sensing measurement setup.

\subsubsection{Sensing Session Termination}
In the sensing session termination, STAs stop performing measurements and terminate the sensing session.

\begin{figure*}[htbp]
	\centering
	\includegraphics[width=1\linewidth]{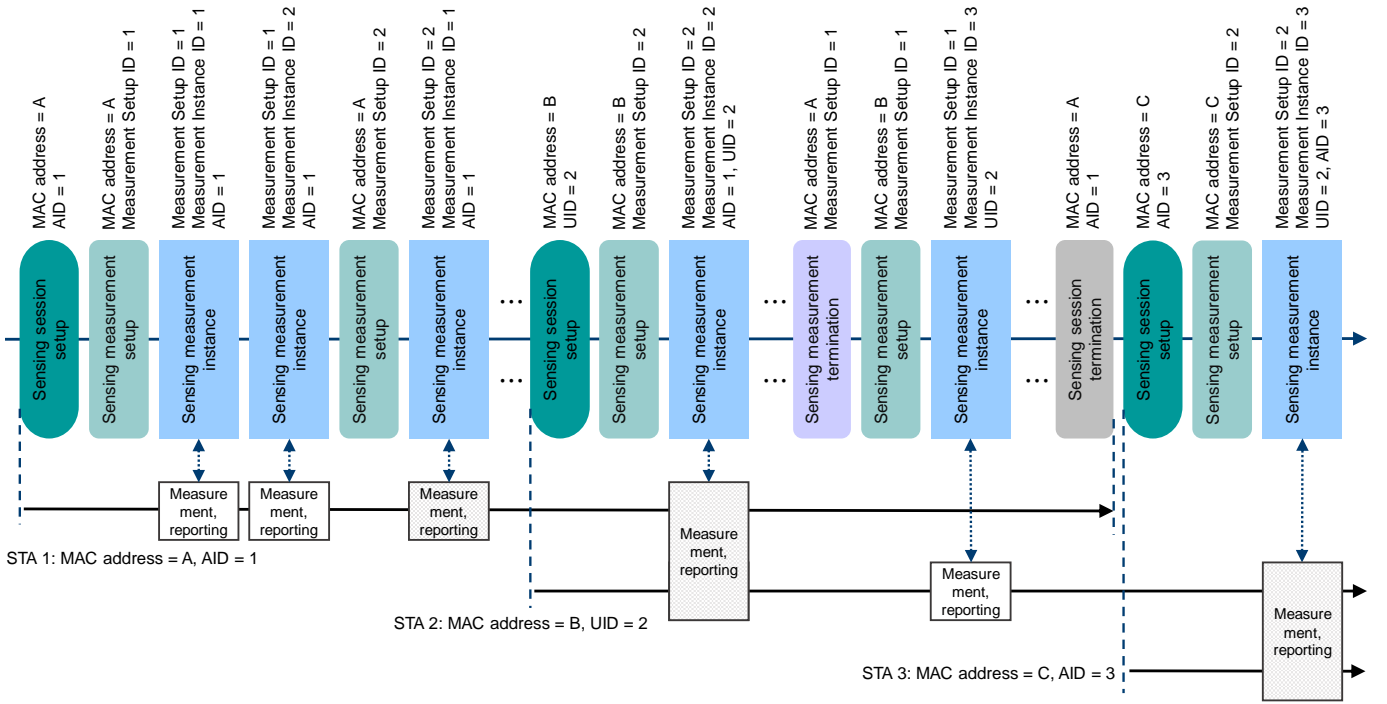}
	\caption{Example of WLAN sensing procedure.}
	\label{11bf_Examp_protocol}
\end{figure*}

An example of the general WLAN sensing procedure is shown in Fig. \ref{11bf_Examp_protocol}, where an AP performs WLAN sensing with three non-AP STAs, which are referred to as STA 1, STA 2, and STA 3 with MAC addresses A, B, and C, respectively. STA 1 has AID 1, STA 2 has UID 2, and STA 3 has AID 3. The example starts with a sensing session setup procedure performed between the AP and STA 1, which establishes a sensing session identified by the AID of STA 1 (AID 1). A first sensing measurement setup procedure is then performed, which defines a set of operational attributes labeled with a Measurement Setup ID equal to 1. After the sensing measurement setup, sensing measurement instances are performed based on the defined operational attribute set (Measurement Setup ID equal to 1). Each measurement instance is labeled with a Measurement Instance ID. After some time, a second sensing measurement setup procedure is performed between the AP and STA 1, which defines a second operational attribute set that is labeled with a Measurement Setup ID of 2. After the second sensing measurement setup, any subsequent sensing measurement instances may be performed based on either the first (Measurement Setup ID equal to 1) or the second (Measurement Setup ID equal to 2) operational attribute sets. An operational attribute set may be terminated by performing a sensing measurement setup termination procedure. For example, Measurement Setup ID equal to 1 is terminated for the sensing session between the AP and STA 1.

\subsection{Sensing Measurement Instance}
Depending on whether there exist trigger frames in the sensing measurement procedure, two broad measurement configurations are considered here, i.e., Trigger-based (TB) configuration and non-TB configuration. Accordingly, a specific sensing measurement instance can be one of the two options, i.e., a TB sensing measurement instance or a non-TB sensing measurement instance.

\begin{figure*}[htbp]
	\centering
	\includegraphics[width=1\linewidth]{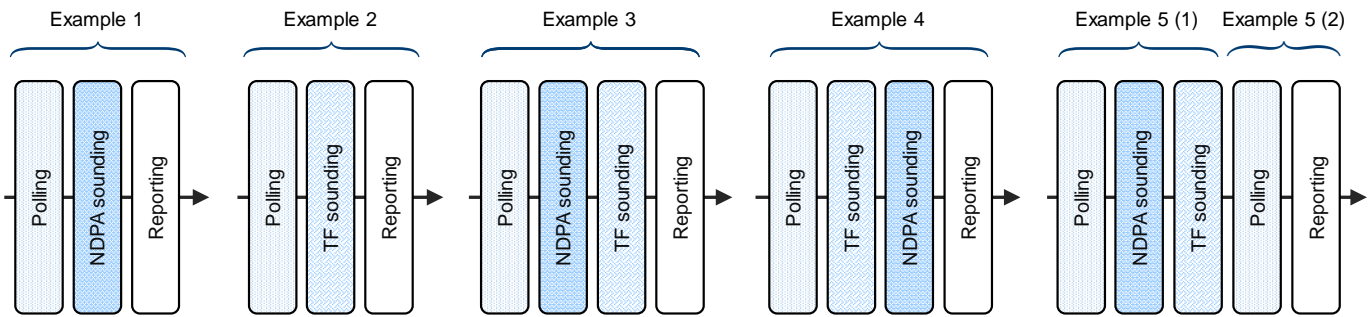}
	\caption{Examples of possible TB sensing measurement instances.}
	\label{TB_instance_comb}
\end{figure*}

\subsubsection{TB sensing measurement instance}
The TB sensing measurement instance is a trigger-based variant of the sensing measurement instance for the case where the AP is the sensing initiator, and one or more non-AP STAs are the sensing responders. It may comprise a polling phase, null data packet (NDP)\footnote{A NDP is a special PPDU that includes a preamble portion but no payload.} announcement (NDPA) sounding phase, trigger frame (TF) sounding phase, and reporting phase. Note that any combination in the order of these phases may be present in the TB sensing measurement instance, as shown in Fig. \ref{TB_instance_comb}. To effectively measure the channel between the initiator and multiple responders, the initiator AP should first perform polling to identify the responder STAs that are expected to participate in the upcoming sensing sounding in the TB sensing measurement instance. If STAs are likely to perform upcoming sensing sounding, they can return a response to request participation in a TB sensing measurement instance. Polling should always be performed to check the availability of the responder STA before performing the actual sensing measurement in the TB sensing measurement instance. After the polling for sensing, the initiator AP can then perform sensing measurement with the responder STAs. In the NDPA sounding phase, the initiator AP, which is a sensing transmitter, sends an NDP to the STAs that are sensing receivers and that have responded in the polling phase to perform downlink sensing sounding. In the TF sounding phase, the initiator AP, as a sensing receiver, requests the responder STAs, which are sensing transmitters and respond in the polling phase, to perform NDP transmission for uplink sensing sounding. It is noticed that both NDPA sounding phase and TF sounding phase are optionally present, and will only be present if at least one responder STA that is a sensing receiver/transmitter has responded in the polling phase. The last phase of a TB sensing measurement instance is the reporting phase. 

\begin{figure*}[htbp]
	\centering
	\includegraphics[width=1\linewidth]{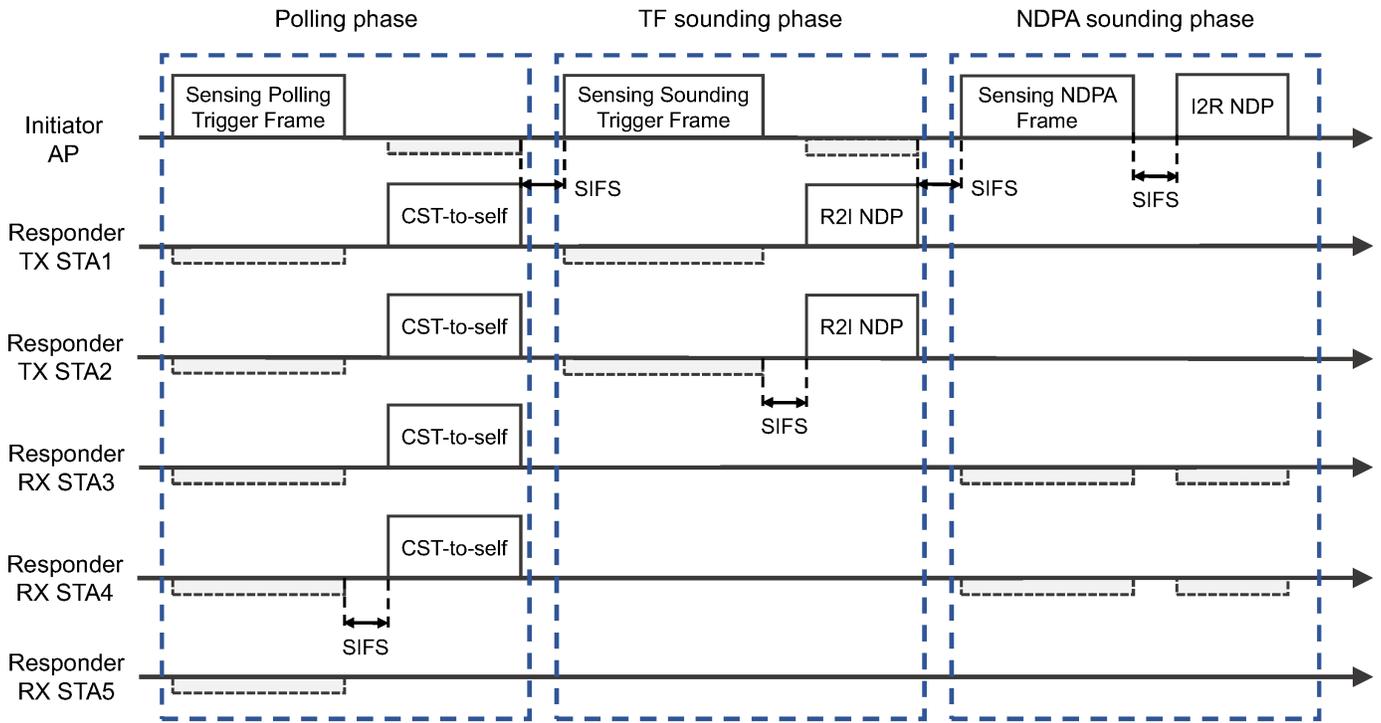}
	\caption{Example of TB sensing measurement instance.}
	\label{TB_instance}
\end{figure*}

Fig. \ref{TB_instance} shows an example of a TB sensing measurement instance consisting of a polling phase, an NDPA sounding phase, and a TF sounding phase. In the polling phase, the AP polls five STAs, where STA 1 and STA 2 are sensing transmitters and STA 3, STA 4, and STA 5 are sensing receivers. STA 1-STA 4 return responses (e.g., Clear to send (CTS)-to-self) to the AP, so both a TF sounding phase and NDPA sounding phase are present. In the TF sounding phase, the AP sends a Sensing Sounding Trigger frame to STA1 and STA 2 to solicit Responder-to-Initiator (R2I) NDP transmissions. In the NDPA sounding phase, the AP sends a Sensing NDP Announcement frame followed by Initiator-to-Responder (I2R) NDP to STA3 and STA 4. There is a short interframe space (SIFS) between each frame.

In the reporting phase, sensing measurement results are reported, and the corresponding sensing measurement reporting can be either immediate or delayed. During the reporting phase, the transmitter AP sends a trigger frame to the receiver STAs to request sensing measurement results obtained from the I2R NDP of the current measurement instance when an immediate feedback reporting is provided, or from the I2R NDP of the previous measurement instance when a delayed feedback reporting is provided. For the delayed reporting, a responder STA can send delayed measurement reports for multiple sensing measurement setups together as a single feedback. 

\begin{figure*}[htbp]
	\centering
	\includegraphics[width=1\linewidth]{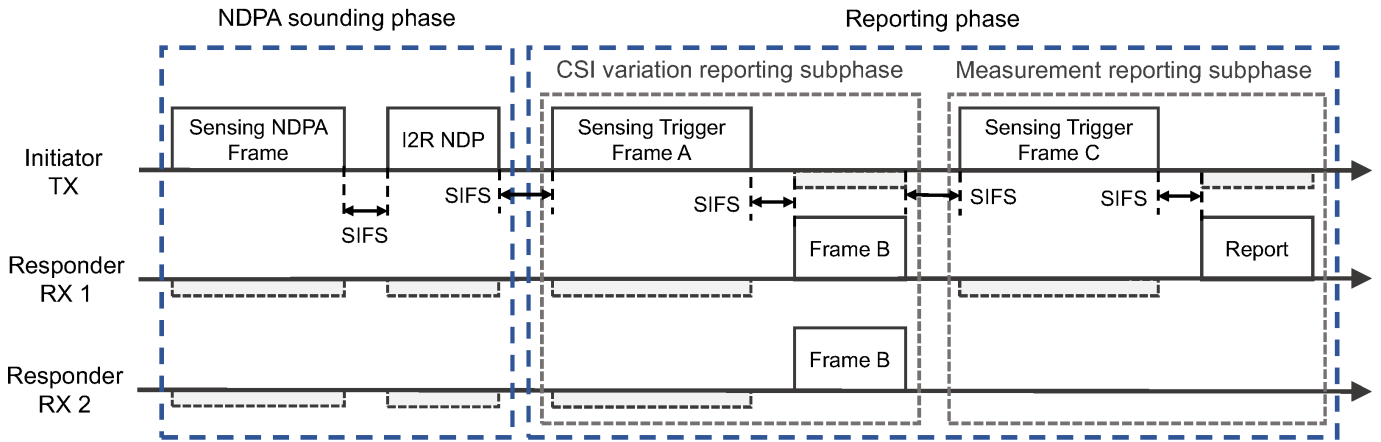}
	\caption{Example of threshold-based reporting phase in a TB sensing measurement instance.}
	\label{Threshold_reporting}
\end{figure*}

Moreover, IEEE 802.11bf also provides an optional threshold-based reporting phase in the TB sensing measurement instance. It is applicable in the case where the sensing initiator of the TB sensing measurement instances is a sensing transmitter. The optional threshold-based reporting phase consists of a CSI variation reporting subphase and possibly a measurement reporting subphase. The CSI variation represents the quantified difference between the currently measured CSI and the previously measured CSI at the sensing receiver. The CSI variation threshold to be compared to the CSI variation value for each sensing responder, is determined by the sensing initiator, and different sensing responders have different thresholds. In the CSI variation reporting subphase, after receiving the trigger frame from the sensing initiator, the sensing responder shall send the CSI variation feedback value (in frame B) to the initiator for threshold comparison. In the measurement reporting subphase, only sensing responders with CSI change values greater than or equal to the CSI change threshold assigned to them are required to provide feedback on the measurement results. An example of the threshold-based reporting phase in a TB sensing measurement instance is shown in Fig. \ref{Threshold_reporting}.

\begin{figure}[htbp]
	\centering
	\includegraphics[width=1\linewidth]{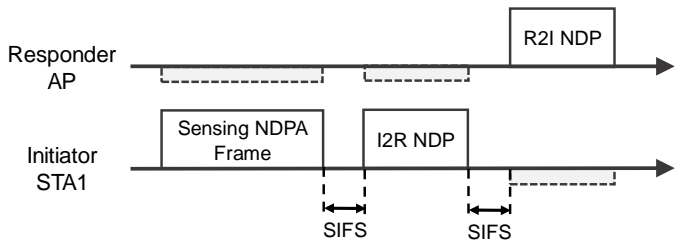}
	\caption{Example of non-TB sensing measurement instance.}
	\label{non_TB_instance}
\end{figure}

\subsubsection{Non-TB sensing measurement instance}
The non-TB sensing measurement instance is a non-trigger-based variant of the sensing measurement instance for the case where a non-AP STA is the sensing initiator, and an AP is the sensing responder. In the non-TB sensing measurement instance, when the initiator STA is both a sensing transmitter and a sensing receiver, it first sends a Sensing NDPA frame to the responder AP to configure the parameters for the subsequent I2R NDP and R2I NDP. Then, after going through an SIFS, an I2R NDP is sent to perform uplink sensing sounding. Upon correct reception of the NDPA frame, the responder AP should send an R2I NDP as a response to the initiator STA to perform downlink sensing sounding. Sensing feedback for the I2R sounding can be sent to the initiator STA after the R2I NDP transmission. When the initiator STA is the sensing transmitter, sensing detection is performed only during I2R NDP transmission, while the R2I NDP is continued to be sent by the responder AP as an acknowledgment of the receipt of the sensing NDP frame and the I2R NDP. In this case, the R2I NDP will be transmitted at the minimum possible length. When the initiator STA is the sensing receiver, R2I NDP is responsible for the sensing detection while I2R NDP is transmitted at the minimum possible length to maintain a unified flow. More details can be found in \cite{11bf_NonTB_instan}. An example of the non-TB-based sensing measurement instance is shown in Fig. \ref{non_TB_instance}.

\subsection{DMG Sensing Procedure}
\begin{figure}[htbp]
	\centering
	\includegraphics[width=1\linewidth]{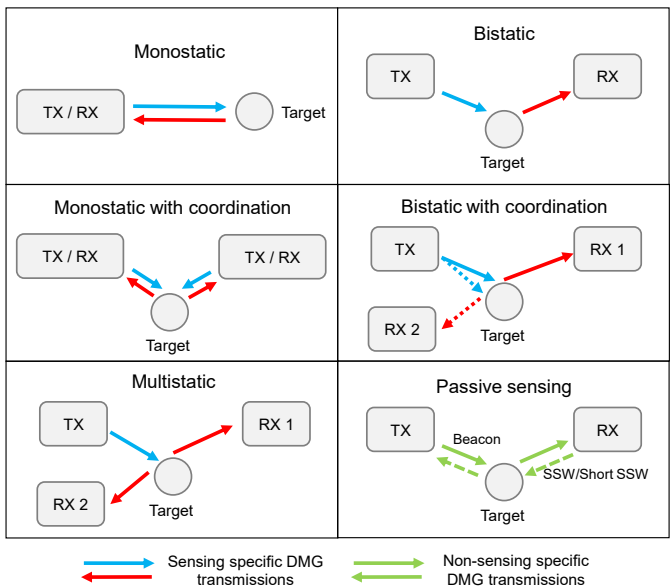}
	\caption{Examples of different DMG sensing types.}
	\label{DMG_sensing_taxonomy}
\end{figure}

IEEE 802.11ad and IEEE 802.11ay make full use of beamforming at the 60 GHz band in the (E)DMG implementations to compensate for the severe path loss, and provide multi-Gigabit-per-second data-rate throughput. IEEE 802.11bf extends these techniques to cope with the challenges of sensing in the mmWave band, and designs an efficient sensing procedure, denoted as DMG sensing procedure. The DMG sensing procedure is an improved version of the general WLAN sensing procedure to support highly directional sensing in the 60 GHz band. As opposed to sub-7 GHz sensing (WLAN sensing), DMG sensing offers wider channel bandwidth and smaller wavelength (allowing the use of compact antenna arrays for beamforming), thus enabling higher range resolution and angular resolution. Depending on the number and roles of the devices involved in sensing, there are various types of DMG sensing, including monostatic, bistatic, multistatic, monostatic sensing with coordination, bistatic sensing with coordination, and passive sensing, as shown in Fig. \ref{DMG_sensing_taxonomy}. Different DMG sensing types have different sensing processes, which will be described in detail later. By incorporating innovative sensing techniques and procedures in the frequency band range around 60 GHz, IEEE 802.11bf enables more accurate sensing applications.

\begin{figure}[htbp]
	\centering
	\includegraphics[width=1\linewidth]{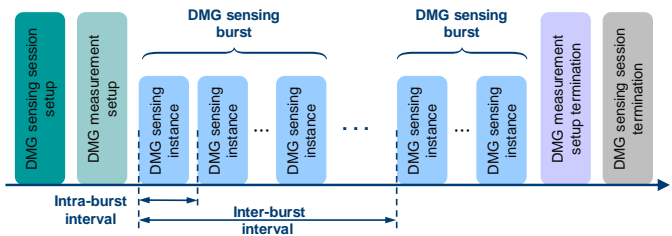}
	\caption{Example of DMG sensing procedure.}
	\label{DMG_sensing_procedure}
\end{figure}

As shown in Fig. \ref{DMG_sensing_procedure}, a DMG sensing procedure generally comprises a DMG sensing session setup, DMG measurement setup, DMG sensing burst, DMG sensing instance, DMG measurement setup termination, and DMG sensing session termination. In particular, a DMG sensing burst is a virtual concept that defines a set of multiple DMG sensing instances in order to perform Doppler estimation in each burst. IEEE 802.11bf specifies the time between consecutive instances in a DMG sensing burst as the intra-burst interval, and the time between consecutive bursts as the inter-burst interval. Note that the DMG sensing procedure is a subset of the WLAN sensing procedure, so the rules for WLAN sensing procedure also apply to DMG sensing procedure unless otherwise stated.

Prior to the DMG sensing procedure, it is assumed that beamforming training between the sensing initiator and the sensing responder(s) is completed in advance, which facilitates the exchange of preamble, data, and synchronization information between them in the DMG sensing procedure. At the beginning of the DMG sensing procedure, DMG sensing capabilities are exchanged between the sensing initiator and the sensing responder to identify the type of DMG sensing. Then, a set of operational attributes associated with DMG sensing bursts and DMG sensing instances are defined in the DMG measurement setup, which may include intra-burst and inter-burst schedule, roles of the sensing initiator and sensing responder, and other parameters. After the setup of the DMG sensing procedure, DMG sensing instances are performed based on the defined operational attribute set to perform channel measurements. A DMG sensing instance typically contains three phases, i.e., initiation phase, sounding phase, and reporting phase. It is worth noting that only the sounding phase is mandatory, while the initiation and reporting phases are optional. According to different DMG sensing types, the specific implementation of DMG measurement instances will be different. More details are given as follows.

\subsubsection{Monostatic Sensing}
In monostatic sensing, any IEEE 802.11-compatible PPDU suitable for sensing can be transmitted by a monostatic (E)DMG STA. Since the sensing transmitter and sensing receiver are the same STA, there is no need for interoperability with any uninvolved STAs. When an uninvolved STA receives a standard compliant signal transmitted by a monostatic (E)DMG STA, the PPDU may be measured at the PHY layer, but will be discarded at the MAC layer because the frame is not addressed to it.

\begin{figure*}[htbp]
	\centering
	\subfigure[Bistatic DMG sensing instance in which the sensing initiator is the sensing transmitter.]{
		\includegraphics[width=0.75\linewidth]{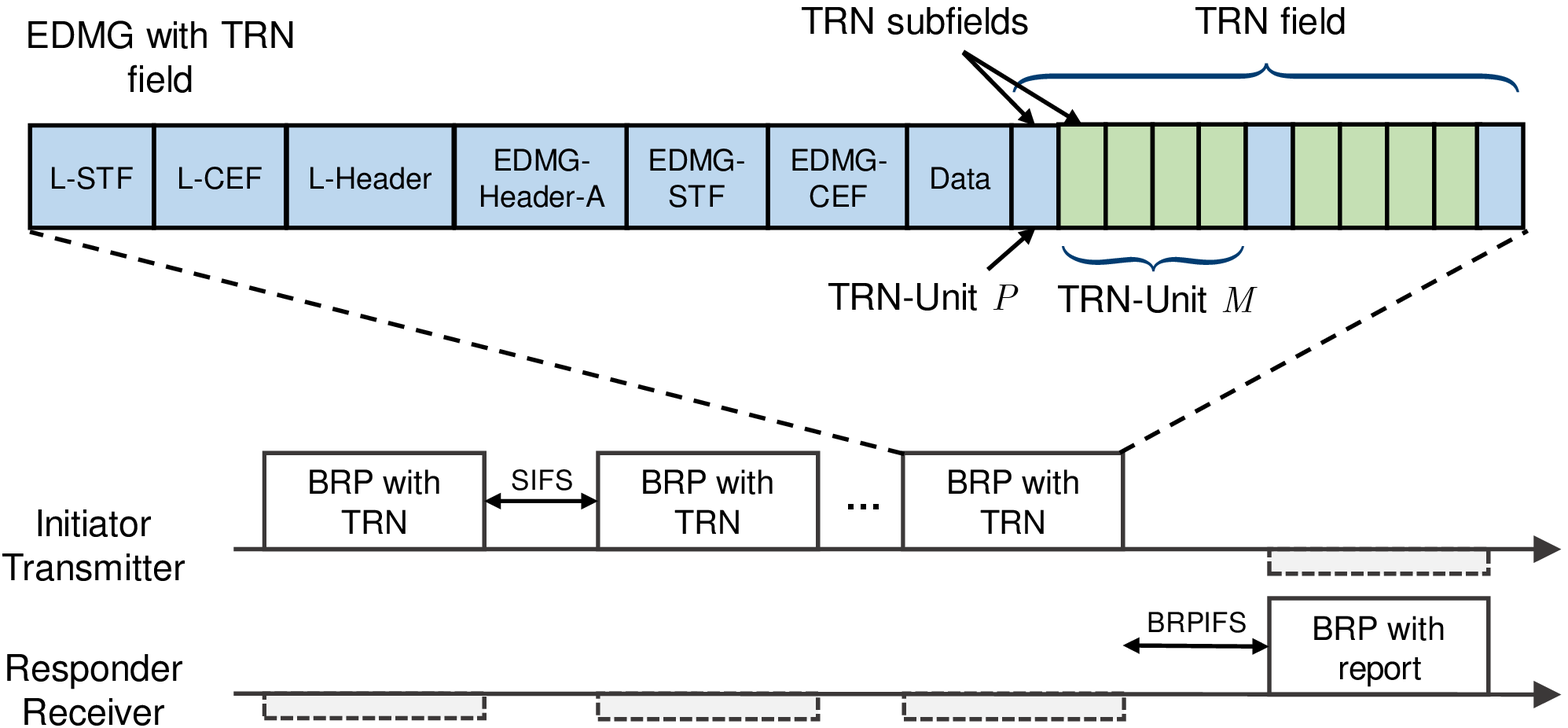}
		\label{DMG_bistatic_instance:1}
	}
	\\
	\subfigure[Bistatic DMG sensing instance in which the sensing initiator is the sensing receiver.]{
		\includegraphics[width=0.75\linewidth]{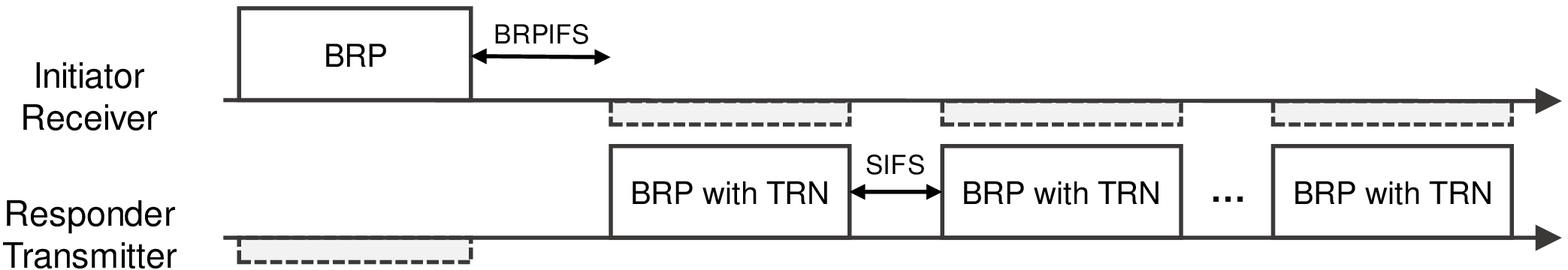}
		\label{DMG_bistatic_instance:2}
	}
	\caption{Examples of DMG sensing instance for bistatic sensing.}
	\label{DMG_bistatic_instance}
\end{figure*}

\subsubsection{Bistatic Sensing}
In bistatic sensing, the sensing transmitter and the sensing receiver are two distinct STAs. The bistatic DMG sensing instance consists of a number of Beam Refinement Phase (BRP) frames with a training (TRN) field transmitted by the sensing transmitter and one BRP frame sent by the sensing receiver, as shown in Fig. \ref{DMG_bistatic_instance}. IEEE 802.11bf reuses the IEEE 802.11ay beamforming training (BFT) method during a DMG sensing instance by using the TRN field appended to the end of the BRP frames. The TRN field is composed of a series of TRN subfields. The first $ P $ TRN subfields are responsible for synchronization and channel estimation, where the sensing transmitter uses the same antenna weight vector (AWV) configuration towards the sensing receiver as the data field. Note that the AWV is a vector of weights imposed on each element of an antenna array that enables the energy of the beam to be concentrated in a narrow range and emitted in a certain direction, i.e., beamforming. In the next $ M $ TRN subfields, the AWV may be changed in each TRN subfield to sweep through all the beams to cover the sensing environment. The initiator shall choose the format of the TRN field in each of the transmitted BRP frames in a way that it is compatible with the responder capabilities, such as BRP-TX, BRP-RX, or BRP-RX/TX PPDUs. For the case where the sensing initiator is the sensing transmitter, the initiator transmits the BRP frames with the TRN field during the sounding phase, and the responder receives these frames and performs the sensing measurements on the TRN fields. In the reporting phase, the responder responds with a BRP frame containing the report as channel measurement feedback. For the case where the sensing initiator is the sensing receiver, the initiator first transmits a BRP frame. Then, BRP frames with a TRN field are transmitted by the responder. There is no reporting phase in the receive initiator DMG bistatic measurement instance since the sensing initiator is the sensing receiver. In particular, if the DMG sensing subfield in the BRP request field is equal to 1, it indicates that the PPDU of the current BRP frame is for sensing and will not be used for beamforming training. At this point, the BRP frames sent by the sensing initiator will contain the BRP Sensing element in which the control and management parameters necessary for initiating the sensing instance are set.

\begin{figure*}[htbp]
	\centering
	\includegraphics[width=1\linewidth]{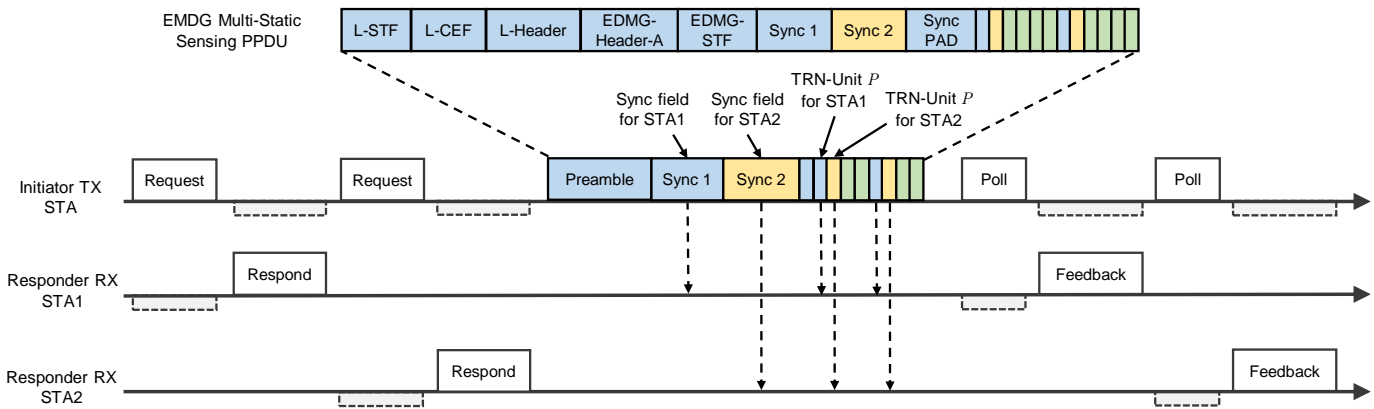}
	\caption{Example of DMG multistatic sensing instance.}
	\label{Multistatic_sensing_instance}
\end{figure*}

\subsubsection{Multistatic Sensing}
In multistatic sensing, the sensing transmitter and more than one sensing receivers are distinct STAs. Multistatic sensing is significantly different from bistatic sensing because the preamble/data field of the existing PPDU is only for a specific STA, while other STAs may not receive it, if the PPDUs are transmitted with the same AWV as the data field. Furthermore, multiple devices need to be synchronized to a single time base. Thus, an initiation phase of multistatic EDMG sensing instance is needed to prepare the different responder STAs for the measurement, and to schedule the synchronization and the TRN field. To enable multiple receivers to participate in sounding using the same PPDU (i.e., EMDG Multi-Static Sensing PPDU), IEEE 802.11bf allows multiple synchronization (Sync) fields to be inserted after the EDMG-STF (if present) or the EMDG-Header-A to replace the EDMG-CEF and data fields of the PPDU to ensure synchronization of the different receivers. A padding field ensures that the length of the Sync fields together is equal to a multiple of the TRN-Unit length. The TRN field of an EDMG Multi-Static Sensing PPDU is identical to the TRN field of an EDMG PPDU, with the exception that multiple TRN-Unit P subfields are transmitted with the AWV to the respective receivers. Fig. \ref{Multistatic_sensing_instance} illustrates a DMG sensing instance of multistatic sensing. The handshake between the initiator and the responders activates the responders to be ready to participate in the sounding and report in sequence during the reporting phase. After receiving the response from the last responder, the initiator transmits the EMDG Multi-Static Sensing PPDUs for synchronization and DMG sensing purposes. In particular, the sensing responders STA 1 and STA 2 are synchronized with the initiator using the corresponding Sync fields. In the reporting phase, the sensing initiator polls each of the responders for a sensing report, and the sensing responders respond in the predefined order.

\subsubsection{Monostatic Sensing with Coordination}
Monostatic sensing with coordination is an extension of monostatic sensing, where the transmissions of one or more devices performing monostatic sensing is coordinated by a sensing initiator. The sensing initiator with the coordinated monostatic sensing can be a STA involved in sensing measurements or a STA that does not have monostatic sensing capability. Similarly to multistatic sensing, there is a need to coordinate multiple monostatic devices, so scheduling and control information is exchanged between the sensing initiator and all sensing responders during the initiation phase. In coordinated monostatic sensing mode, the sensing initiator may request the sensing responders to transmit and receive a monostatic PPDU in a specific direction by indicating the TX/RX beams to be used in each measurement burst. IEEE 802.11bf specifies that the sounding for each responder can be performed either sequentially or simultaneously. In monostatic sensing with coordination, each collaborating STA only uses its own clock to transmit and receive sensing signals, so there is no need to synchronize between different STAs as in multistatic sensing. This greatly reduces the synchronization overhead.

\subsubsection{Bistatic Sensing with Coordination}
Bistatic sensing with coordination is an extension of the bistatic sensing to coordinate multiple sensing responders by one sensing initiator. Unlike coordinated monostatic sensing, the sounding phases of the different responders are sequential in the sounding phase of coordinated bistatic sensing, where the sounding procedure for each responder is the same as that of bistatic sensing.

\subsubsection{Passive Sensing}
In passive sensing, transmissions that are not specifically designed for sensing are used by other devices for sensing, such as beacon frames, sector sweep (SSW) frames, and short SSW frames. IEEE 802.11bf has provided an efficient downlink DMG passive sensing method based on the beacon frames in the beacon transmission interval (BTI) of a beacon interval (BI). Note that in (E)DMG, beacons are periodically transmitted in BTI to many directions (sectors) for network announcements and initiator transmit sector sweep (I-TXSS) at the AP. Support for passive sensing in the beacon is optional and is indicated by the Passive Sensing Support subfield that is set to 1 in the Short Sensing Capability element transmitted in the beacon frames. During the transmission of the beacon frames, non-AP STAs can find the sector ID of a beacon that provides the highest signal quality. In IEEE 802.11ay, non-AP STAs do not need to know the specific transmit direction (sector) corresponding to the sector ID of a beacon with the highest signal quality, but only need to feed it back to the AP. However, in the passive sensing of IEEE 802.11bf, STAs that are interested in sensing need to know in which direction the beacon was transmitted and the location information of the AP to interpret the obtained sensing. To this end, STAs request sensing information by sending an Information Request frame to the AP, while the AP responds with an Information Response frame that contains the beacon directions and the AP location.

In \cite{11bf_passive_ABFT}, an uplink DMG passive sensing method based on association beamforming training (A-BFT) of BI is proposed. This differs from BTI, in that it enables an AP to perform I-TXSS, A-BFT following the BTI performs responder transmit sector sweep (R-TXSS) at the STAs by using SSW frames or Short SSW frames. Based on this, A-BFT can be used by an AP for passive sensing. Similar to the procedure of the beacon based passive sensing method for downlink, an AP that wants to sense the environment uses the received SSW frames or Short SSW frames to perform passive sensing. To obtain sensing information about the direction of each sector of SSW/Short SSW frames and the STA location, the AP sends an Information Request frame to the non-AP STA, and the non-AP STA returns the required information by using the Information Response frame. Note that the AP can obtain passive sensing results from multiple non-AP STAs to provide a broader sensing area.

\subsection{Sensing by Proxy Procedure}
The sensing by proxy (SBP) procedure allows a non-AP STA (SBP initiator) to request an AP (SBP responder) to perform WLAN sensing on its behalf. In particular, the SBP initiator STA may obtain sensing measurements of the channel between an AP and one or more non-AP STAs in the SBP. To establish an SBP procedure, the SBP initiator shall first send an SBP Request frame to an SBP capable AP as a proxy. After accepting the SBP procedure request, the AP performs a WLAN sensing procedure with one or more non-AP STAs. In this stage, the STA as SBP initiator can participate as a sensing responder in this WLAN sensing procedure to enlarge the sensing area. Finally, the AP reports the obtained sensing measurements back to the SBP initiator that requested them. With the execution of the SBP procedure, it is possible for a non-AP STA to obtain necessary sensing measurements for detecting and tracking changes in the environment.

\begin{figure}[htbp]
	\centering
	\includegraphics[width=1\linewidth]{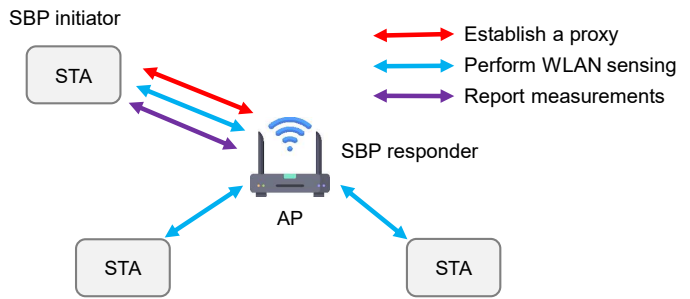}
	\caption{Illustration of sensing by proxy.}
	\label{SBP}
\end{figure}

\section{IEEE 802.11bf: Candidate Technical Features}
To enable and enhance sensing functionality, a variety of candidate technical features have been proposed by numerous industrial and academic experts during IEEE 802.11bf meetings. In the following, we focus on technologies that attracted the most attention.

\begin{figure*}[ht]
	\centering
	\subfigure[]{
		\includegraphics[width=.3\linewidth]{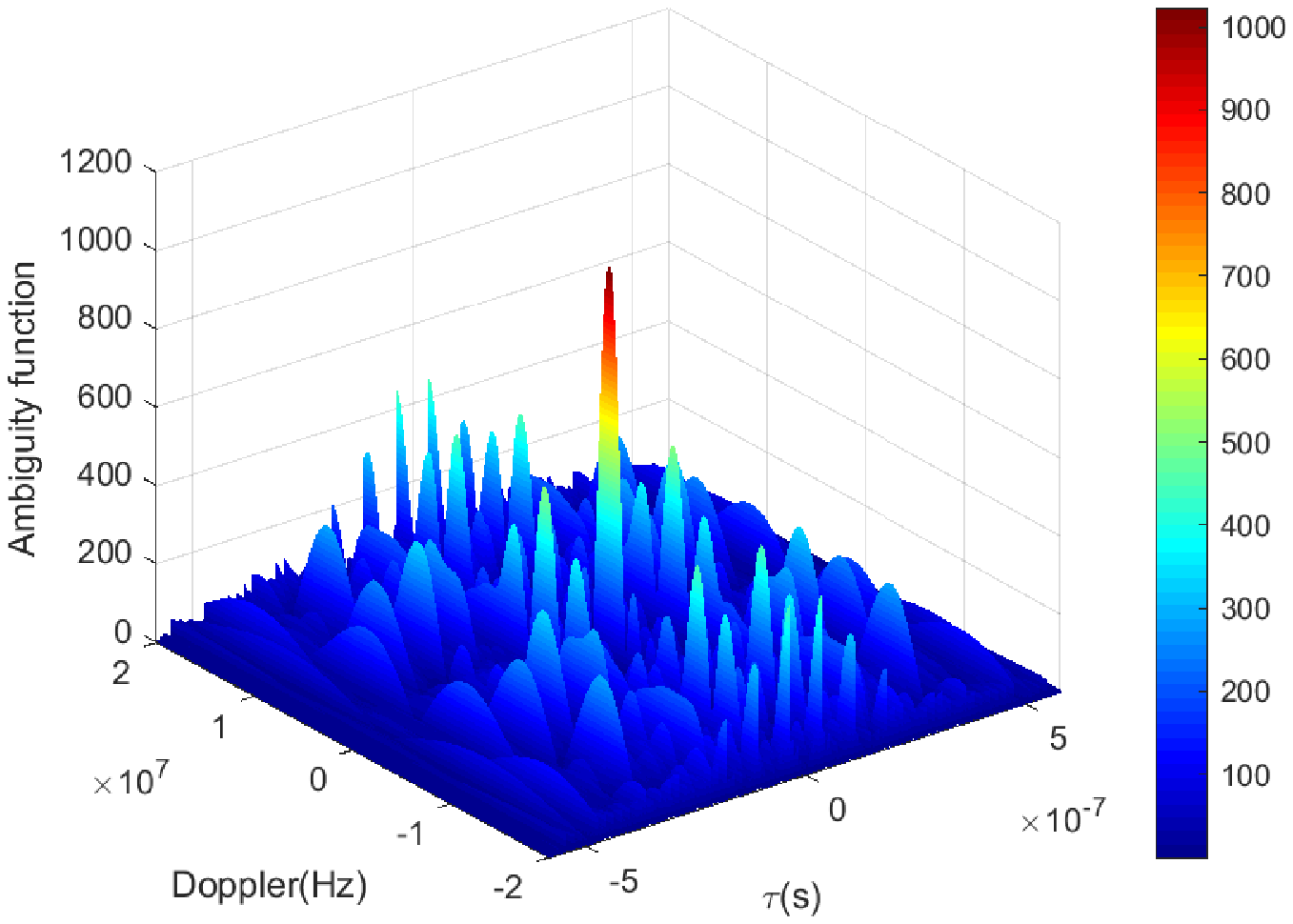}
		\label{demo_waveform(a)}
	}
	\subfigure[]{
		\includegraphics[width=.3\linewidth]{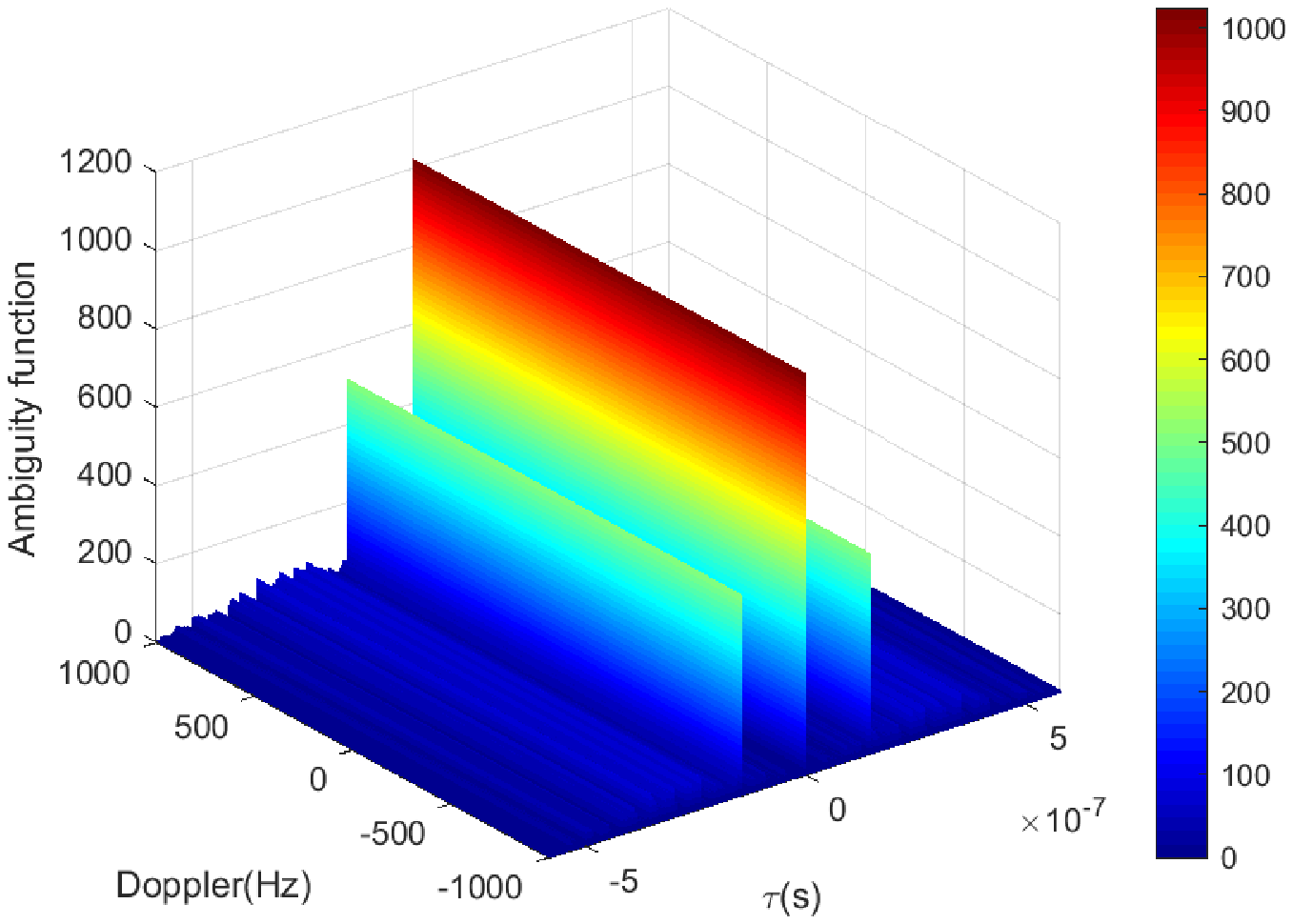}
		\label{demo_waveform(b)}
	}
	\subfigure[]{
		\includegraphics[width=.3\linewidth]{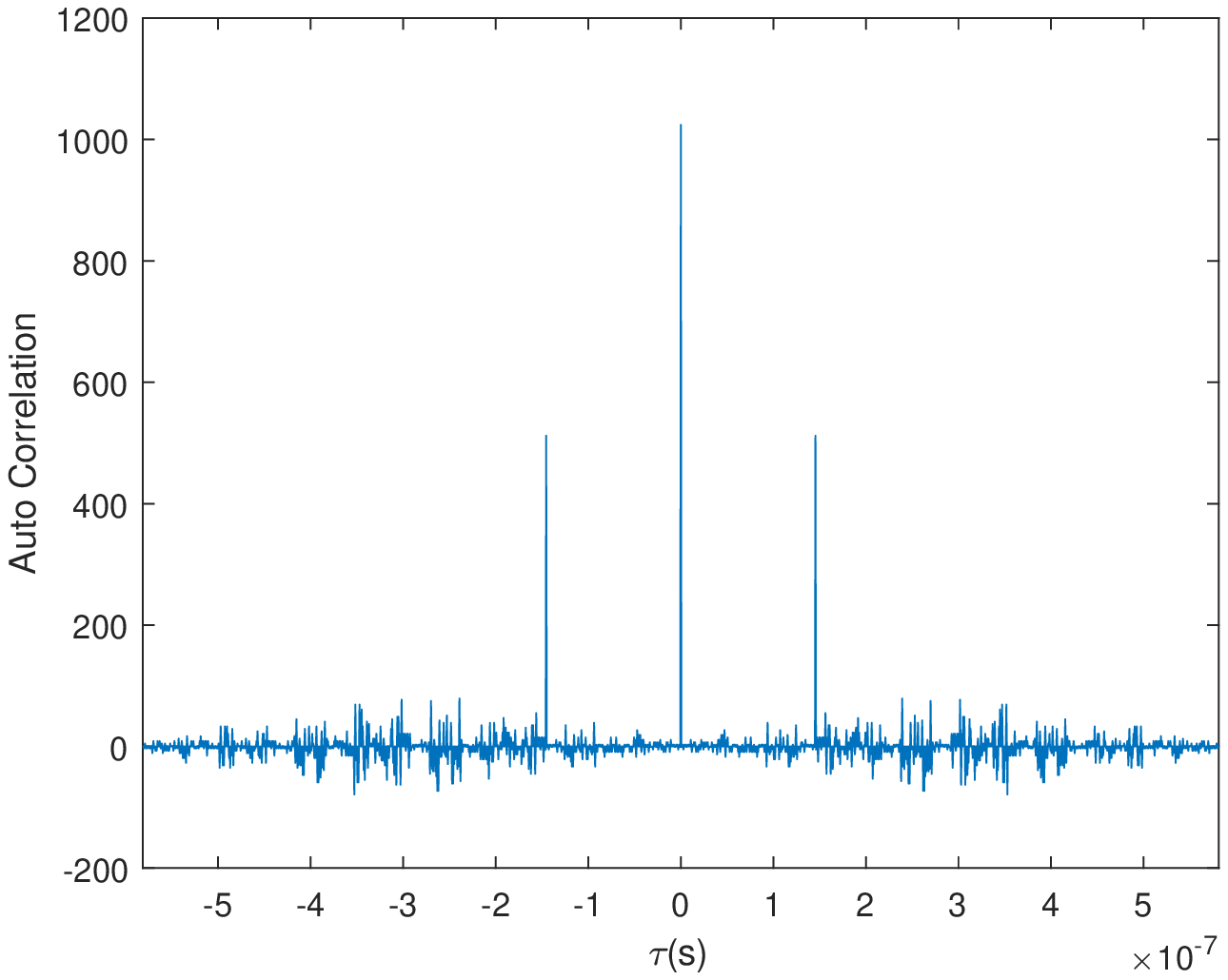}
		\label{demo_waveform(c)}
	}
	\\
	\subfigure[]{
		\includegraphics[width=.3\linewidth]{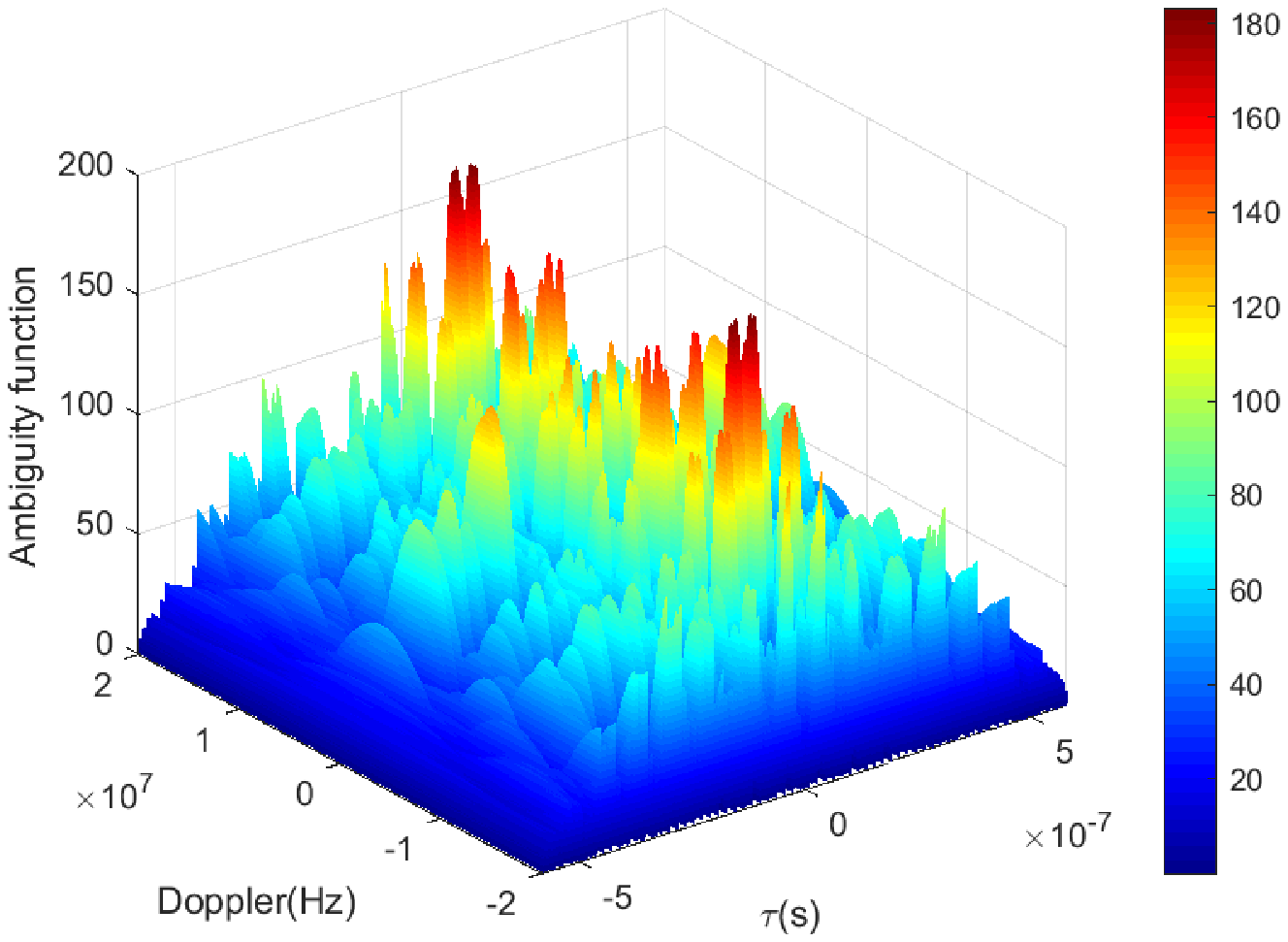}
		\label{demo_waveform(d)}
	}
	\subfigure[]{
		\includegraphics[width=.3\linewidth]{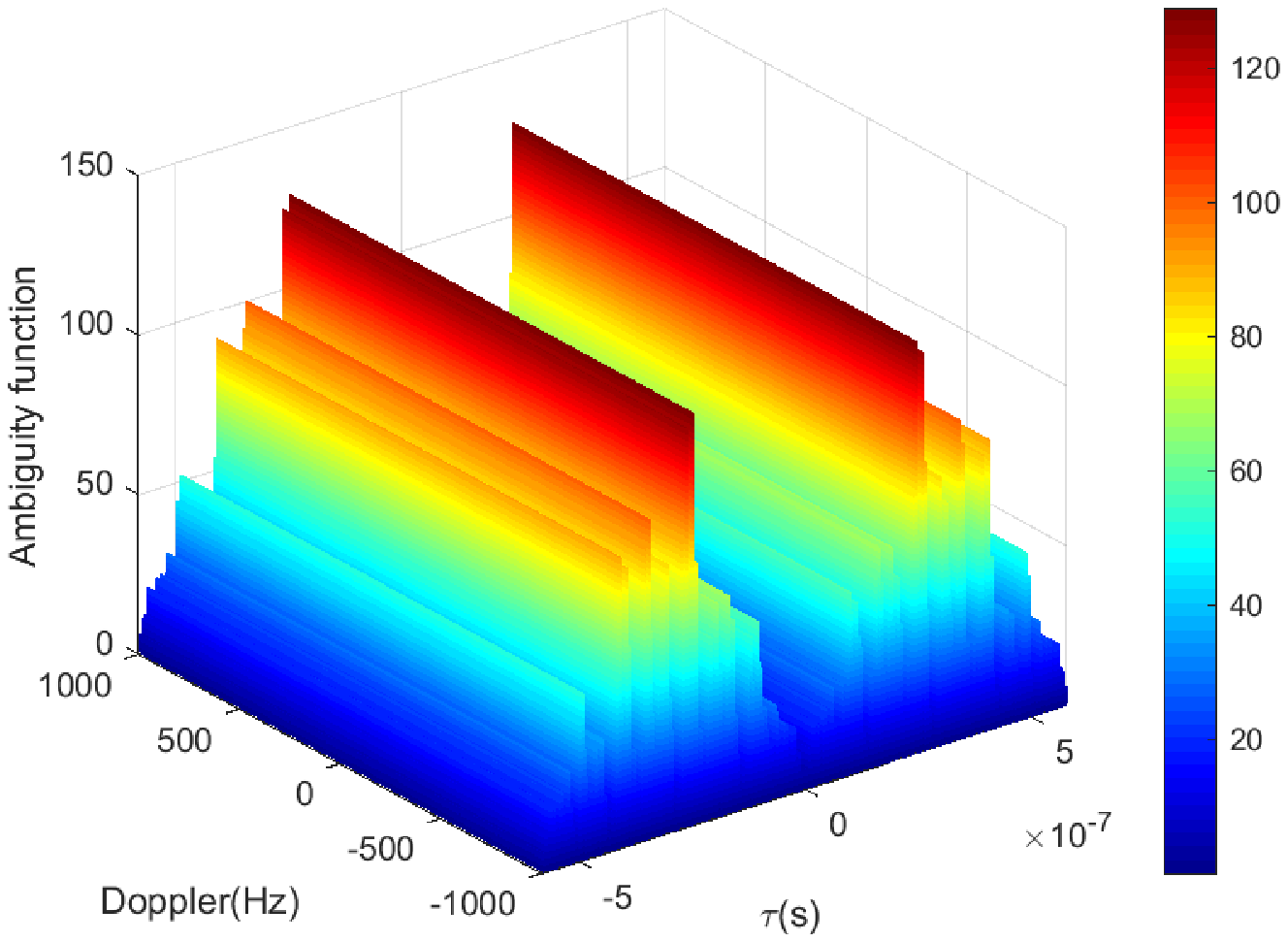}
		\label{demo_waveform(e)}
	}
	\subfigure[]{
		\includegraphics[width=.3\linewidth]{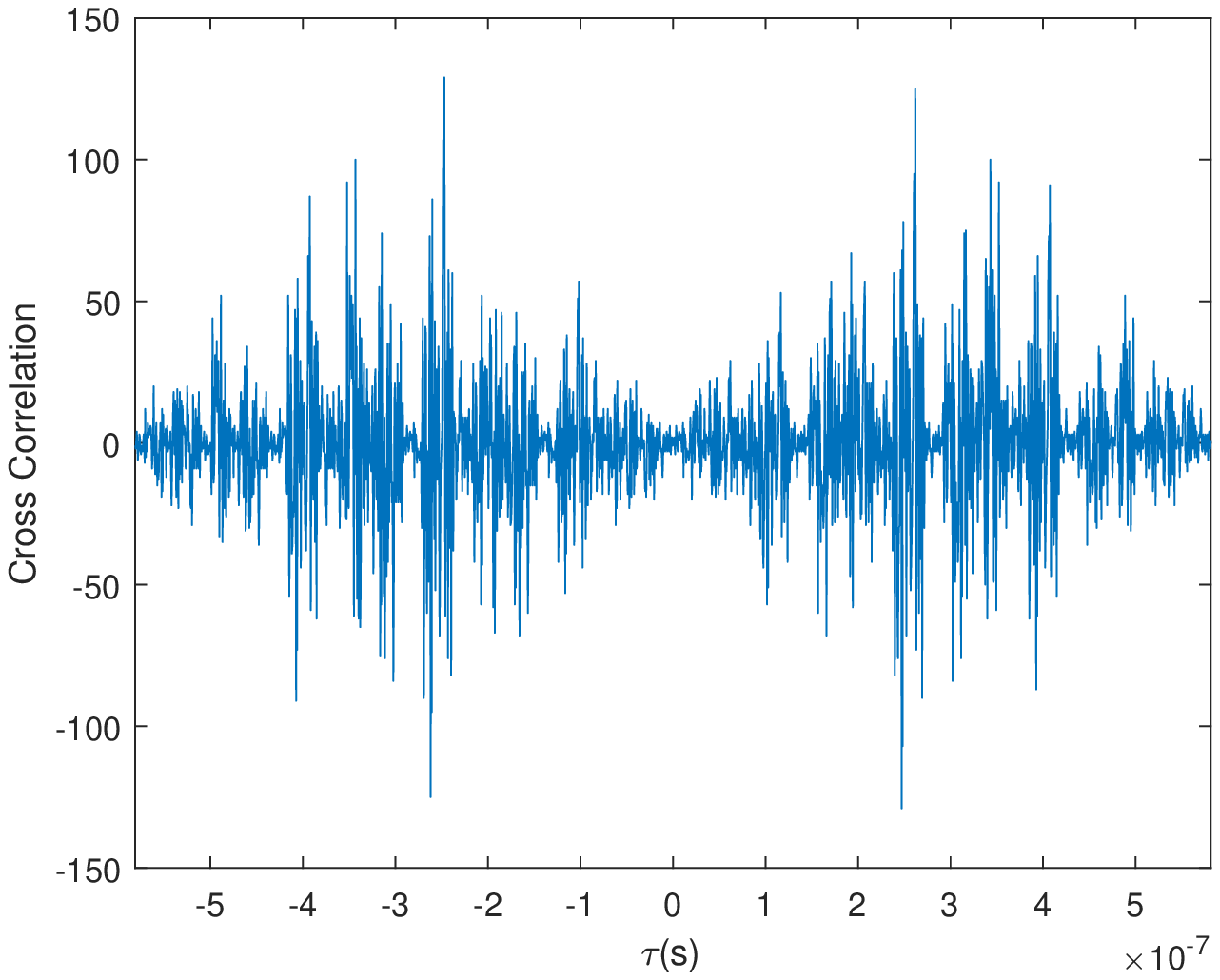}
		\label{demo_waveform(f)}
	}
	\caption{A demo AAF for CE0 in the range of (a) $[-20\text{MHz},20\text{MHz}]$, (b) $[-1000\text{Hz},1000\text{Hz}]$; (c) Auto Correlation result.
		A demo CAF for CE0 and CE1 in the range of (d) $[-20\text{MHz},20\text{MHz}]$, (e) $[-1000\text{Hz},1000\text{Hz}]$; (f) Cross Correlation result.}
	\label{demo_waveform}
\end{figure*}

\begin{figure*}[htbp]
	\centering
	\includegraphics[width=1\linewidth]{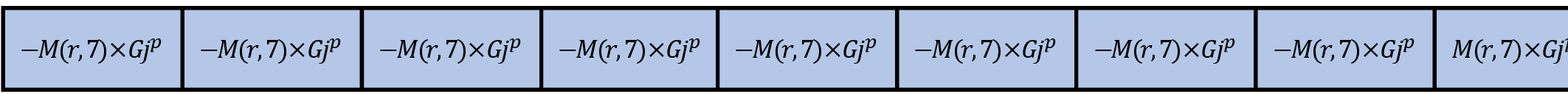}
	\caption{Sync subfield structure.}
	\label{fig:SyncSubfield}
\end{figure*}

\begin{figure}[htbp]
	\centering
	\includegraphics[width=0.7\linewidth]{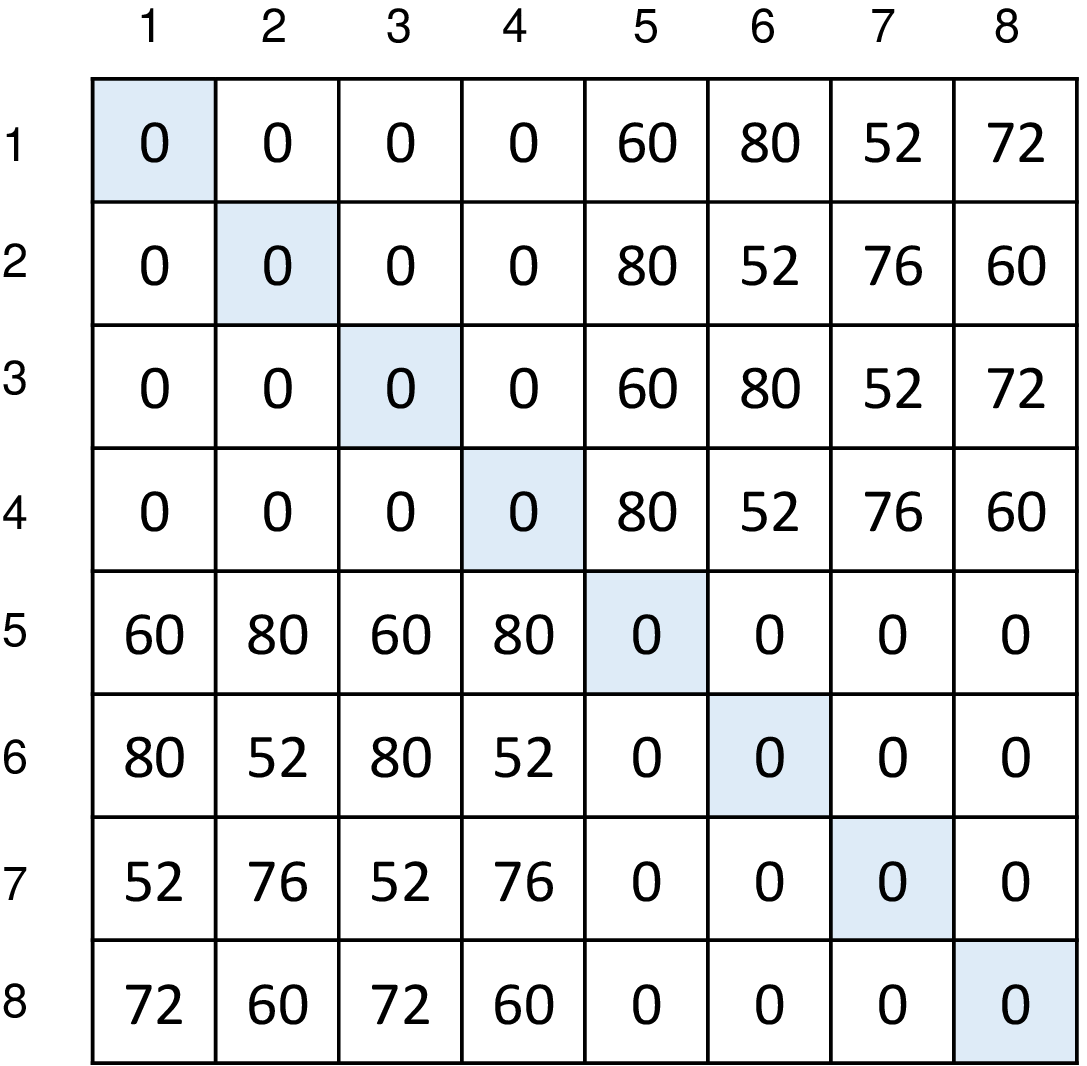}
	\caption{Correlation of 8 sequences.}
	\label{fig:SequencePerfor}
\end{figure}

\subsection{Waveform/Sequence Design}
The existing preamble waveform/sequences (e.g., Sync sequences) were specifically designed for communication systems, focusing on enhancing the communication performance (e.g., PAPR: Peak to Average Power Ratio). However, the properties of sensing (e.g., angle/range resolution, Doppler frequency) were not considered. In other words, new waveform/sequence and waveform evaluation metrics are needed in WLAN sensing.

In order to take sensing performance into account, one of the most straightforward ideas is to use the ambiguity function (AF) to evaluate the transmit waveforms \cite{11bf_sequence}. The ambiguity function is one of the most widely used tools in radar waveform analysis, which processes the received signal through a matched filter and shows a R-D (Range-Doppler) map result. It is defined as \cite{Ambiguity}:
\begin{align}
	|\mathcal{X}_{a,b}(\tau,f_d)| = \left|\int_{-\infty}^{\infty}S_a(t)S_b^*(t-\tau)e^{j2\pi f_d t}dt\right|^2,
\end{align}
where $S_a(t)$ and $S_b(t)$ denote two sequences to be matched, and $\tau$ and $f_d$ denote time delay (range) and Doppler frequency, respectively. $|\mathcal{X}_{a,b}(\tau,f_d)|$ is called the auto ambiguity function (AAF) if $S_a(t)=S_b(t)$, and the cross ambiguity function (CAF) if $S_a(t) \neq S_b(t)$. A demo of magnitudes of the AAF and CAF outputs is shown in Fig.~\ref{demo_waveform}\ref{sub@demo_waveform(a)}\ref{sub@demo_waveform(d)}, using a bandwidth setup of $20$ MHz and sequences 
\begin{align}
	CE0 = [Ga^7, -Gb^7, Ga^7, -Gb^7, Ga^7, Gb^7, Ga^7, Gb^7], \label{CE0} \\
	CE1 = [Ga^8, -Gb^8, Ga^8, -Gb^8, Ga^8, Gb^8, Ga^8, Gb^8] \label{CE1}
\end{align}
with 128-length Golay sequences $Ga^7$, $Gb^8$, as defined in IEEE 802.11ay \cite{802.11ay}.

A good AAF should look like a ``thumbtack'' shape that has a peak at the origin and low/zero side lobes elsewhere, while a good CAF should have as small as possible magnitudes everywhere. It can be clearly observed that AAF and CAF usually do not have a good performance at the whole bandwidth, as shown in Fig.~\ref{demo_waveform}\ref{sub@demo_waveform(a)}\ref{sub@demo_waveform(d)}. However, considering the Doppler frequency in actual scenarios (e.g., a living room), it is much lower than the signal bandwidth, meaning that an AF design can be narrowed down to a much smaller area around the origin. Driven by this, a new concept named LAZ/ZAZ (Low/Zero Ambiguity Zone), and its corresponding signal designs are presented in \cite{LAZ/ZAZ}, where LAZ/ZAZ can be considered as a local AF limited by a maximum Doppler frequency and maximum time delay. Fig.~\ref{demo_waveform}\ref{sub@demo_waveform(b)}\ref{sub@demo_waveform(e)} shows an example for local AF with maximum Doppler frequency $f_{max} = f_c*v_{max}/c = \frac{60 \text{GHz}*5\text{m/s}}{3*10^8\text{m/s}}=1000\text{Hz}$. It is observed that the local AF has good ``ambiguity'' as the maximum value of the AAF side lobes and maximum value of CAF are relatively very low in the region of interest.

Not only that, but the auto/cross correlation serves as a surrogate function for local AAF/CAF in the case of 60 GHz sensing. Looking back at Fig.~\ref{demo_waveform}\ref{sub@demo_waveform(b)}\ref{sub@demo_waveform(e)}, any magnitudes of AF along Doppler frequency axis (time delay) remain constant, which means that the evaluation of local AF performance can be restricted to a certain Doppler frequency. The auto/cross correlation in Fig.~\ref{demo_waveform}\ref{sub@demo_waveform(c)}\ref{sub@demo_waveform(f)} is actually a zero Doppler frequency of the local AAF/CAF. As such, IEEE 802.11bf currently uses auto/cross correlation to evaluate waveform designs.

By considering the auto/cross correlation performance of the sequences, a new structure of synchronization sequences for Sync subfields in multi-static sensing is proposed \cite{11bf_Sync}, as shown in Fig. \ref{fig:SyncSubfield}, where $r$ denotes the index of the STA. The Sync subfields of different STAs use different rows of the coefficient matrix $M(r,c)$, which is defined as
\begin{align}
	M =
	\begin{bmatrix}
		1 & -1	& 1	& -1 & 1 & 1 & 1 & 1  \\
		1 & -1	& 1	& -1 & 1 & 1 & 1 & 1  \\
		1 & 1 & -1 & -1 & 1 & -1 & -1 & 1  \\
		1 & 1 & -1 & -1 & 1 & -1 & -1 & 1  \\
		-1 & 1 & -1	& 1 & 1 & 1 & 1 & 1  \\
		-1 & 1 & -1	& 1 & 1 & 1 & 1 & 1  \\
		1 & -1 & -1 & 1 & -1 & -1 & 1 & 1  \\
		1 & -1 & -1 & 1 & -1 & -1 & 1 & 1  
	\end{bmatrix}.
\end{align} 
Therefore, the proposed design can support a maximum of 8 different STAs for sensing simultaneously. For $r$ from 1 to 8, $Gi^p$ is $Ga^7, Ga^8, Ga^7, Ga^8, Gb^7, Gb^8, Gb^7, Gb^8$, and $Gj^p$ is $Gb^7, Gb^8, Gb^7, Gb^8, Ga^7, Ga^8, Ga^7, Ga^8$, respectively. The Sync subfields need to be designed to have high auto and low cross correlations. Fig. \ref{fig:SequencePerfor} gives the performance measurement of the synchronization subfield proposed in \cite{11bf_Sync}, where the elements on the diagonal denote the maximum of the sidelobe of the auto-correlation of the same Sync sequence, while the non-diagonal elements denote the maximum magnitude of the cross-correlation of different Sync sequences. Notice that the maximum value of the auto-correlation for Syn sequences is 1024. It is shown that the proposed Sync sequences can ensure that the cross-correlation between the initial or last 4 sequences is 0, while the cross-correlation between the initial 4 and last 4 sequences is kept at a low level.

\subsection{Feedback Types}
In order to implement various sensing services efficiently, it is critical to provide appropriate and accurate measurement results in the feedback. To support this, one or more types of sensing measurement results and their formats need to be defined depending on the requirements of different sensing applications at sub-7 GHz and 60 GHz. 

\begin{figure}[htbp]
	\centering
	\subfigure[TCIR]{
		\includegraphics[width=0.45\linewidth]{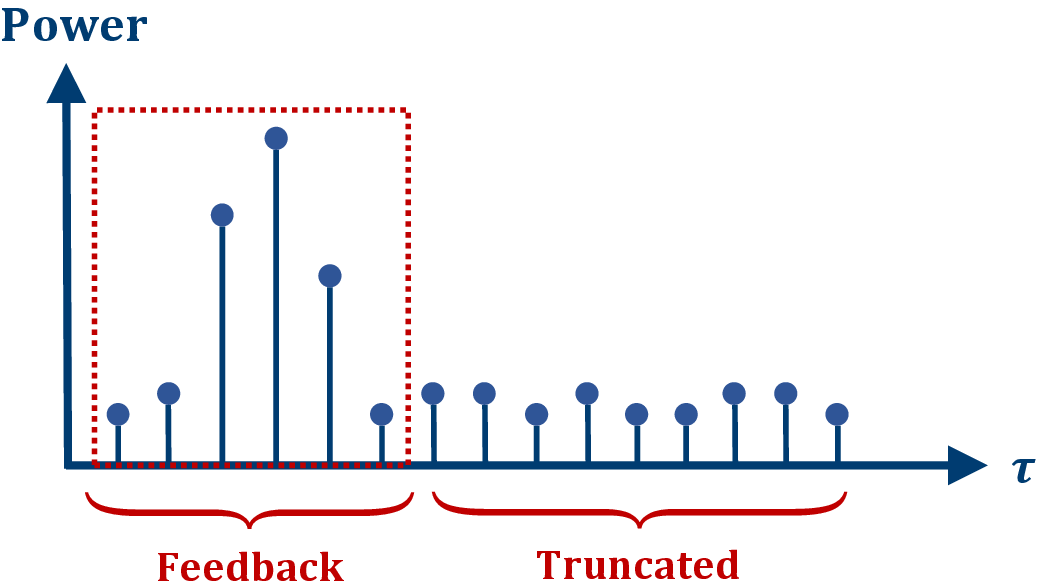}
		\label{fig:TCIR}
	}
	\subfigure[Differential quantization]{
		\includegraphics[width=0.45\linewidth]{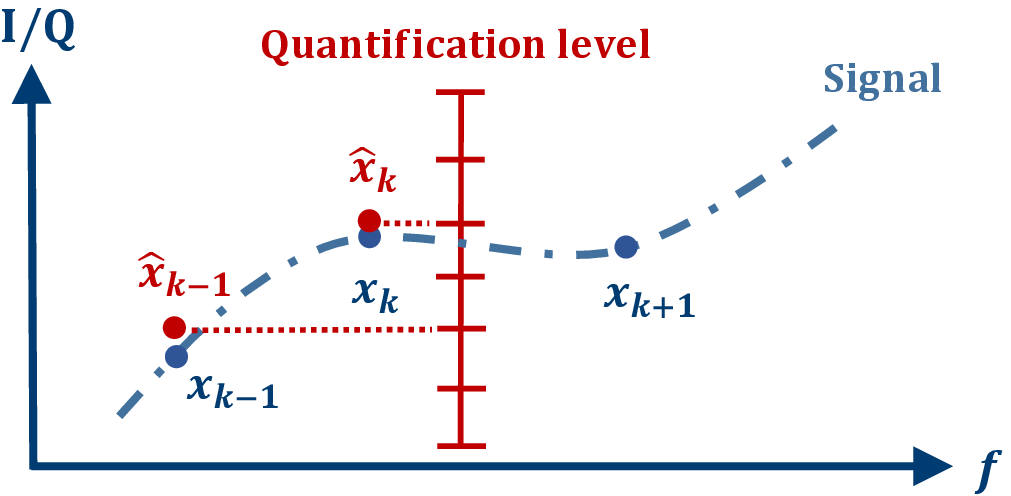}
		\label{fig:DifferQuanti}
	}
	\caption{Potential feedback types for sub-7GHz.}
	\label{fig:fb_sub7}
\end{figure}

For sub-7 GHz sensing, the potential feedback types include the full CSI matrix, partial CSI\cite{11bf_PartialCSI}, truncated power delay profile (TPDP)/truncated channel impulse response (TCIR)\cite{11bf_TCIR}, and frequency-domain differential quantization\cite{11bf_DifferQuanti}, etc. First, it is fairly straightforward to take advantage of measurements that are already defined in the IEEE 802.11 standard, such as those for explicit feedback\footnote{Note that the explicit feedback corresponds to the case when the RX needs to directly send the downlink measurements to the TX, while the implicit feedback corresponds to that the TX uses channel reciprocity to obtain the downlink measurements based on the uplink measurements.}. Thus, the full CSI matrix, a typical measurement/feedback type that has been extensively used in various sensing implementations, is repurposed for IEEE 802.11bf in the sub-7 GHz band. It captures the wireless characteristics of the signal propagation between the transmitter and the receiver at certain carrier frequencies, and thus represents the channel frequency response (CFR). Because CSI is a direct result of the channel estimation, it retains the full environmental information and has no information loss compared to other feedback types. Next, by noting that some sensing applications use only either the amplitude or the phase of the CSI, not both, it is  proposed to reduce the overhead by feeding back either the amplitude or the phase\cite{11bf_PartialCSI}. Another alternative feedback type is TPDP/TCIR. Specifically, \cite{11bf_TCIR} proposed the TPDP/TCIR as a potential WLAN sensing measurement result in explicit feedback, as shown in Fig. \ref{fig:TCIR}. By performing an inverse fast fourier transform (IFFT) on the CSI in the frequency domain, the CIR in the time domain can be calculated. It describes the multipath propagation delay versus the received signal power for each channel. Since the maximum range of WLAN sensing is a few dozen meters \cite{11bf_use_case}, the first few complex values of the CIR already contain the desired environment information and thus can be reported as sensing measurement results. Furthermore, to further reduce the feedback overhead in the frequency domain, a frequency-domain differential quantization was proposed in \cite{11bf_DifferQuanti}, as shown in Fig. \ref{fig:DifferQuanti}. The main idea behind this is to report the differential signal obtained by subtracting the quantized signal of the previous subcarrier from the signal of the current subcarrier. Since the quantization range is reduced by the differential operation, the feedback overhead is reduced while ensuring the integrity of the measurement information. 

\begin{figure}[htbp]
	\centering
	\includegraphics[width=1\linewidth]{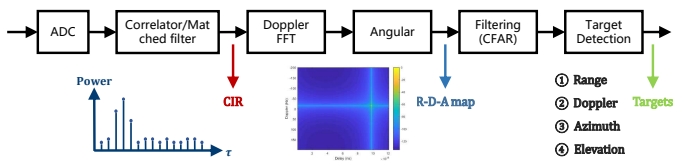}
	\caption{A typical sensing image processing flow diagram \cite{11bf_SensingImage} for 60GHz.}
	\label{fig:SensingImage}
\end{figure}

As for the 60 GHz band, three types of sensing measurement result have been identified for IEEE 802.11bf, i.e., channel measurement for IEEE 802.11ad/ay, Range-Doppler-Angular map (R-D-A map) and target-related parameters. First, similar to TCIR, the channel measurement in IEEE 802.11ad/ay returns the CIR corresponding to each TRN subfield, where the number of returned delay taps is optional. Next, an R-D-A map, also known as sensing image in IEEE 802.11bf, provides an "image" of the surrounding environment, which can be up to a four-dimensional image data consisting of range, Doppler, azimuth, and elevation. Note that by combining some or all of the data from the four dimensions, the sensing image can be a two-dimensional, three-dimensional, or up to four-dimensional image. By detecting the area of higher energy on the R-D-A map, it is possible to know in which location there is a reflector or target, and based on this information subsequent sensing implementation can thus be performed. Furthermore, since direct reporting of R-D-A maps is inefficient and has a high feedback overhead, it is most effective to report target-related parameters, such as the position and Doppler of the target, directly after target detection processing of the R-D-A maps. A typical sensing image processing flow diagram \cite{11bf_SensingImage} for 60 GHz is shown in Fig. \ref{fig:SensingImage}, from which all three types of sensing measurement results can be obtained.

\subsection{Quantization and Compression}
The measurement results need to be quantified before they can be reported to the sensing initiator to reduce the feedback overhead. However, the quantization of measurement results introduces inevitable quantization errors. As a result, designing an efficient and accurate quantization and compression method is critical for WLAN sensing.

The potential quantization and compression methods for IEEE 802.11bf include the legacy quantization procedure in IEEE 802.11n \cite{802.11-2012}, the simplified scaling and quantization method\cite{11bf_SimplScaling}, the power-of-two scaling and quantization method\cite{11bf_LowScaling}, and the fractional scaling and quantization method\cite{11bf_FracScaling}. First, as a proven method, the CSI matrix quantization procedure in IEEE 802.11n \cite{802.11-2012} (or with minor modifications) can be reused for sensing services in IEEE 802.11bf, which is reviewed briefly in the following. Denote $m_H(k)$ as the maximum of the In-phase and Quadrature components of each element of the CSI matrix $\mathbf{H}(k)$ in each subcarrier $k$. Prior to quantization, a scaling factor of the CSI matrix needs to be calculated to improve the dynamic range of quantization. The scaling factor $M_H(k)$ in IEEE 802.11n is obtained by quantizing $m_H(k)$ to three bits in the decibel (dB) domain and then converting back to the linear domain. As a result, the In-phase and Quadrature components of each element in $\mathbf{H}(k)$ are scaled and quantized to $N_b$ bits as
\begin{align}
	h^q(k)={Round}\left(\frac{h(k)}{M_H(k)}\left(2^{(N_b-1)}-1\right)\right), \forall k,
	\label{Eq:RealScaling}
\end{align}
where $h(k)$ denotes the In-phase/Quadrature components of each element of the CSI matrix $\mathbf{H}(k)$, $h^q(k)$ denotes the corresponding quantized data, $N_b$ denotes the number of bits of quantized data, and ${Round}(\cdot)$ rounds a number to its nearest integer. However, the procedure of the CSI matrix quantization in IEEE 802.11n requires considerable computational complexity due to the need for repeated linear to dB and dB to linear conversions. In addition, multiplication and division operations need to be performed prior to quantization.

To avoid these drawbacks, \cite{11bf_SimplScaling} uses a uniform linear scaling factor $M^{lin}_H$ for all subcarriers and simplifies the transformation of $M_H$ from $m_H(k)$ by directly setting $M^{lin}_H$ to the maximum value in $m_H(k),\forall k$, i.e., $M^{lin}_H=\max_k m_H(k)$. The corresponding quantized In-phase and Quadrature components can be obtained by simply replacing $M_H(k)$ in (\ref{Eq:RealScaling}) as $M^{lin}_H$.

Furthermore, \cite{11bf_LowScaling} proposed a low-complexity scaling and quantization method by using the power-of-two scaling factor for the CSI matrix. Let $N_p$ denote the number of bits of original data. To maximize the dynamic range of quantization, the power-of-two scaling factor $\alpha_H$ is chosen to ensure that $2^{(N_p-2)}\le\alpha_H m_H(k)\le 2^{(N_p-1)}-1$. The final quantized In-phase and Quadrature components of each element in $\mathbf{H}(k)$ are given by
\begin{align}
	h^q(k)={Round}\left(\alpha_H h(k)\left(2^{(N_b-N_p)}\right)\right), \forall k.
	\label{Eq:power2Scaling}
\end{align}
Note that since $h(k)$ only needs to be multiplied and divided with powers of two, a binary shift operation can be performed in hardware or software to greatly reduce the computational complexity. Nevertheless, compared to the real-value scaling as in IEEE 802.11n, both the simplified scaling and the power-of-two scaling lead to a relative increase in the quantization error due to the reduction in the dynamic range of quantization.

In addition, considering the trade-off between computational complexity and quantification accuracy, a fractional scaling and quantization method is proposed in \cite{11bf_FracScaling}. Notice that the scaling factor in (\ref{Eq:RealScaling}) can be considered as the ratio between two numbers ($\alpha$ and $\beta$) as 
\begin{align}
	{Round}\left(\frac{2^{(N_b-1)}-1}{m_H(k)}h(k)\right)={Round}\left(\frac{\alpha}{\beta}h(k)\right),
	\label{Eq:RatioScaling}
\end{align}
where $\beta$ can be constrained to be a power of two to reduce the computational complexity and $\alpha$ can be chosen among a set of pre-defined values to optimize the scaling factor. In order to obtain the best performance, the maximum of $\alpha/\beta$ should be found while ensuring the following inequality:
\begin{align}
	\frac{\alpha}{\beta}\le\frac{2^{(N_b-1)}-1}{m_H(k)}.
\end{align}
By expanding the size of the set of $\alpha$, the performance of fractional scaling can approach that of real-value scaling in IEEE 802.11n.

\subsection{Security and Privacy}
With the increase in the number of WLAN devices and the pervasiveness of sensing applications, various technologies using WLAN for location/tracking, recognition/detection, environment mapping, and even through-wall sensing are expected to be widely used in practice. While the development of WLAN sensing has brought us the benefits of an improved life, it has also incurred security and privacy concerns. For instance, due to the broadcast nature of the wireless medium, unauthorized or even malicious entities can use sensing technologies to easily eavesdrop on users' privacy-critical information, such as their daily activities. Even worse, such eavesdropping is often difficult to detect due to the non-intrusive nature of WLAN sensing. As a result, the IEEE 802.11bf amendment not only needs to provide stable sensing services, but also needs to pay more attention to the security and privacy issues. Currently, the IEEE 802.11bf task force is still in discussion about sensing security requirements and the corresponding privacy issues, details of which can be found in \cite{SecurityReq}.

Based on \cite{TG4ab_UBW_Privacy}, the security and privacy issues can be further divided into two categories: sensing report overhearing issues for eavesdropping on the measurement reports in the feedback payload, and sensing packet/signal overhearing issues for eavesdropping on the sensing packets/signals from the legitimate transmitter.

In the sensing report overhearing issues, malicious parties can directly infer the daily activities of a target by collecting the sensing measurement results that are not intended for them. Possible technical solutions include cryptography approaches for secure communication which transmit encrypted channel measurement feedback and physical layer security approaches that apply traditional signal processing methods that result in a significantly lower signal quality received by non-intended users.

In the sensing packet/signal overhearing issues, malicious entities need to estimate the channel measurement results from the eavesdropped sensing packets, and then infer the behavior of the target and thus violate user privacy. In IEEE 802.11az, the legitimate transceiver performs channel estimation by generating and pre-exchanging a protected LTF sequence instead of the pre-defined legacy one. Since the transmitter uses a new LTF sequence, the eavesdropper will estimate the wrong channel measurement results, thus avoiding the information leakage. Furthermore, \cite{CSI_Fuzzer} intentionally manipulates the transmitted signal by imposing an artificial channel response to the signal before transmission. Only the intended user who knows the artificial response can estimate the correct channel measurement and perform sensing.

In order for the correct sensing results to be available only to the sensing initiator, \cite{11bf_Encrypted_TRN} proposes an approach to address the privacy requirements of sensing at 60 GHz using an encrypted sensing measurement. Specifically, a sensing initiator transmits a masked TRN sequence (i.e., TRN1). Based on the legacy TRN sequence (i.e., TRN2), the sensing responder and eavesdropper will estimate a false channel measurement, which includes the information of the actual channel measurement and the difference between TRN1 and TRN2. After obtaining the false channel measurement through feedback, the sensing initiator can correct it back to the actual channel measurement according to the information of TRN1 and TRN2. It is worth noticing that using this proposed approach, the real channel measurements are not available to the sensing responder and eavesdropper, but only to the sensing initiator, thus greatly improving the security. Although there are already various ways to improve security and privacy, IEEE 802.11bf is still discussing which solution to add, not excluding the need for new and innovative solutions, which will require more study.


\section{Simulation and channel model}
In order to evaluate different proposals, simulation and channel models have been discussed in IEEE 802.11bf. Accordingly, the evaluation methodology\cite{11bf_evaluation} and channel model document\cite{11bf_Ch3}, have been developed to facilitate the development of the draft amendment.

\subsection{Evaluation Methodology}
To more objectively and effectively promote the standardization process in IEEE 802.11bf, 
different technical proposals can be evaluated according to the evaluation methodology document \cite{11bf_evaluation}, where PHY performance, a limited set of simulation scenarios, and parameters that might be used when evaluating the performance of different contributions are summarized. 

As yet, two indoor scenarios with dense multi-paths have been proposed (i.e., an indoor living room and conference room) for the simulation, where the room size and the properties of various environmental objects are all strictly defined by the IEEE 802.11bf. During the simulation, one TX-RX pair and a moving target are assumed, and two types of antennas (directional and isotropic) can be adopted. In particular, the TX and RX communicate with each other in a single-input single-output (SISO) mode since the desired antenna pattern can be added to the isotropic antenna. Moreover, in order to simulate the above scenarios, traffic models and hardware impairments are required. As specified in \cite{SENS_evaluation}, they can be the same as that of IEEE 802.11ax for sub-7 GHz \cite{11ax_evaluation} and IEEE 802.11ay for 60 Ghz \cite{11ay_evaluation}.

Furthermore, evaluation criteria between the different contributions need to be defined for IEEE 802.11bf in order to assess the merits of each contribution. As opposed to previous IEEE 802.11 task groups that focused on the communication criteria (i.e., throughput and PER), the IEEE 802.11bf task group adopted estimated parameters accuracy as a metric to evaluate the PHY performance of WLAN sensing. In general, for parameter estimation, the accuracy describes how close the estimated parameter of the intended target (e.g., range, velocity, angle, etc) are to the ground truth\footnote{Ground truth is information provided by direct observation and measurement that is known to be true, rather than by inference.}, and it is described by the root mean square error (RMSE) of the parameters to be estimated, as defined below:
\begin{align}
	\text{RMSE}=\sqrt{\frac{1}{N}\sum_{n=1}^N\left(\hat{x}_n-x_n\right)^2}
\end{align}
where $\hat{x}_n$ is the estimated parameter, $x_n$ is the ground truth, and $N$ is the number of observations. According to \cite{11bf_evaluation}, two comparison criteria of the PHY performance based on the accuracy of estimated parameters, are: a {\it accuracy vs. SNR curves} and a {\it histogram of accuracy}. These are adopted for different operation modes in different scenarios. The accuracy vs. SNR curves indicate the WLAN sensing estimation accuracy with different SNRs. Usually, accuracy improves as SNR increases. As for the histogram of accuracy, the SNRs caused by a moving target on a certain predefined trajectory are different during the WLAN sensing simulation, which results in different estimation accuracies. Based on the estimated accuracy information (e.g., range, velocity, and angle) for each sample, a histogram of accuracy during the simulation can be calculated. 

\subsection{Channel Model}
The initial idea of channel modeling for a WLAN Sensing system was presented in \cite{11bf_Ch1}, where some initial thinking about the goals and necessity of the new channel modeling for WLAN sensing were discussed. Compared with previous channel models operating at sub-7 GHz (e.g., IEEE 802.11ax \cite{11ax_Ch}) and 60 GHz (e.g., IEEE 802.11ay \cite{11ay_Ch}) that focused only on the communication characteristics in specific frequency bands, the new channel model for WLAN sensing should mainly focus on providing accurate space-time characteristics of the propagation channel arising from the device and free moving targets both in the sub-7 GHz and 60 GHz bands. For this reason, two preliminary channel models aiming at WLAN Sensing were established in the channel model document \cite{11bf_Ch3}, each of which is discussed briefly below. 

\begin{figure}[htbp]
	\centering
	\includegraphics[width=1\linewidth]{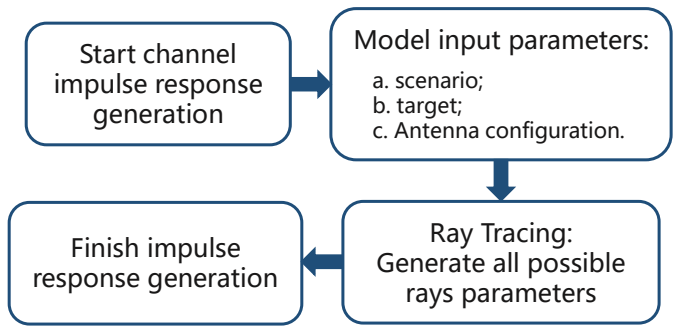}
	\caption{Process of ray-tracing-based channel realization.}
	\label{Ray-Tracing Process}
\end{figure}

The first proposed channel model (known as the ray-tracing-based channel model) is a deterministic model that uses a ray-tracing technique to process channel realization. Ray-tracing technique is a numerical computational electromagnetics method that uses computer program to provide estimates for multipath parameters, e.g., path loss, angle of arrival/departure (AOA/AOD), and time delays, by assuming the transmitted signal as a particle\cite{Ray-Trac}. As illustrated in Fig. \ref{Ray-Tracing Process}, the ray-tracing-based channel model selects scenario, target, and antenna configurations as inputs, and then generates all possible time-variant rays for each TX-RX pair in different simulation frames. Given these, a time-variant channel impulse response (CIR) is generated. For example, in \cite{11bf_Ch3}, two simulation schemes with different configurations in a living room scenario with a device and a free moving target are presented, in which an AP (TX) and STA (RX) are in a monostatic setting and are equipped with a directional/isotropic antennas. The simulation results show the power delay profile (PDP) at different snapshots.

\begin{figure}[htbp]
	\centering
	\includegraphics[width=1\linewidth]{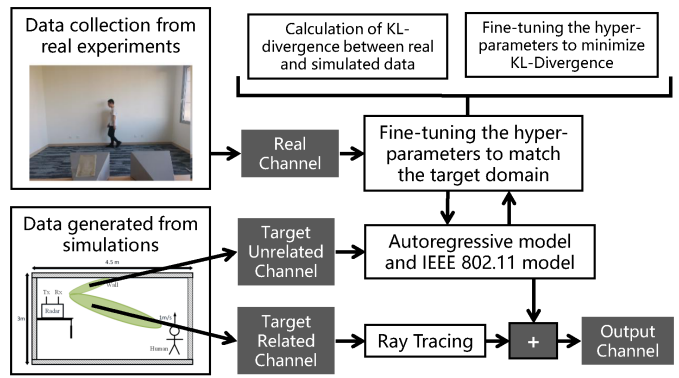}
	\caption{Process of DDHC-based channel realization.}
	\label{fig_DDHC}
\end{figure}

However, ray-tracing-based methods suffer from large computational complexity and can not represent the sensing uncertainty in real measurements such as the unpredictable motions of other objects, and random perturbations of scatter (e.g., diffracted and scattering rays), which are the unrelated rays to those of the target. In order to solve these drawbacks, the data-driven hybrid channel model (DDHC) was also proposed in \cite{11bf_Ch3,11bf_DDHC}. In this so-called hybrid channel model (see Fig.\ref{fig_DDHC} as an example), rays between TX-RX pairs are divided into target-related rays and target-unrelated rays\cite{11bf_DDHC,DDHC}, where the target related-rays are generated via the aforementioned ray-tracing tool and the target-unrelated rays are generated by adopting an autoregressive model of existing standardized channel models (IEEE 802.11ax for sub-7 GHz \cite{11ax_Ch} and IEEE 802.11ay for 60 GHz \cite{11ay_Ch}). On the other hand, real datasets collected from the experiments serve to refine the parameter tuning by minimizing the Kullback-Leibler (KL) divergence between the real and simulated datasets\cite{11bf_DDHC,DDHC}.

\begin{figure}[htbp]
	\centering
	\includegraphics[width=1\linewidth]{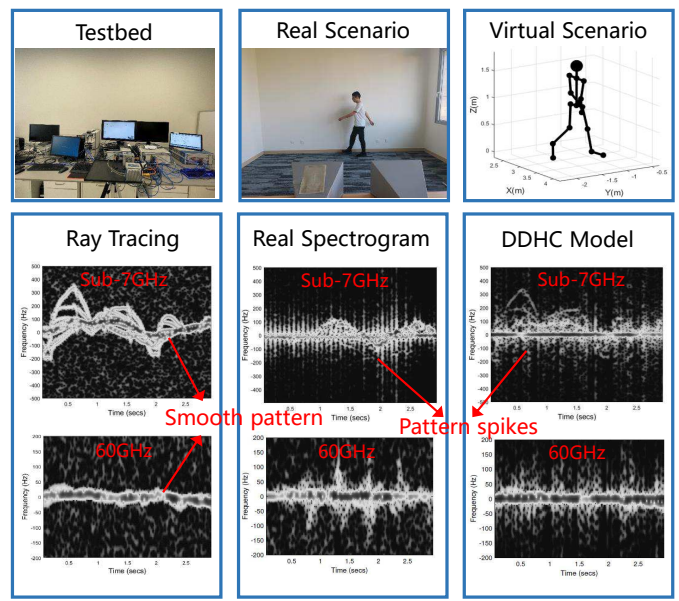}
	\caption{Spectrograms of human motion at sub-7GHz and 60GHz.}
	\label{fig_Spectrograms}
\end{figure}

Fig. \ref{fig_Spectrograms} shows spectrograms of human motion at both sub-7 GHz and 60 GHz, where the conference room scenario with a monostatic transceiver is considered and a person is walking around the room. The comparison results indicate that the ray-tracing-based datasets differ from both the DDHC-based and the real-world datasets in that the motion patterns of the DDHC-based and real spectrogram images have peaks, while the ray-tracing-based ones are smoother. This is due to the unpredictable reflections from the wall and random scatter as well as the irregular movements of the sensing target in the real world. Furthermore, it is verified that the sensing performance of the DDHC model is close to that of a real world dataset and significantly outperforms the performance of the ray-tracing model \cite{11bf_Ch3}.

\begin{figure*}[htbp]
	\centering
	\includegraphics[width=1\linewidth]{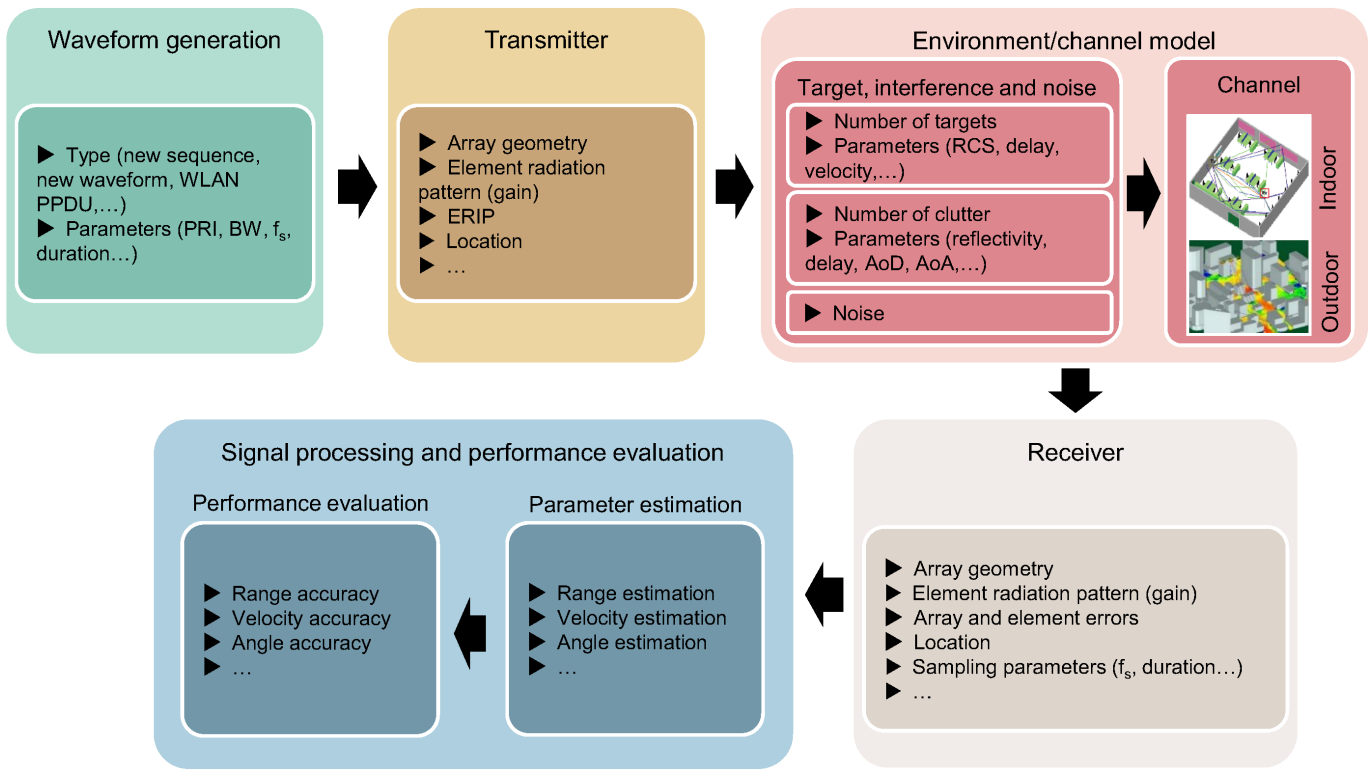}
	\caption{Framework of link level simulation platform.}
	\label{11bf_LLS}
\end{figure*}

\subsection{Link-Level Simulation}
Based on the authenticity and objectivity of the performance verification, IEEE 802.11bf task group pointed out that it is necessary to build an IEEE 802.11bf simulation platform to perform WLAN sensing simulations, and the corresponding link level simulation was proposed \cite{11bf_lls}. The system structure of the link-level simulation platform for WLAN sensing is shown in Fig. \ref{11bf_LLS}, and the specific flow of the simulation is illustrated. 

It is clear that the WLAN sensing link-level simulation mainly contains a waveform generation module, transmitter module, environment/channel module, receiver module, and signal processing and performance evaluation module. Specifically, the waveform generation module is mainly responsible for the generation of different waveforms, which consists of new sequences, new waveforms, normal WLAN PPDU, and modified WLAN PPDUs, etc. The transmitter/receiver module focuses on the transmission/reception of the signal based on some array information, including array geometry, element radiation pattern, and location, etc. Then, the environment/channel module needs to generate the channel matrices according to the transceiver antenna configurations and the parameters of the targets and multi-path defined in \cite{11bf_evaluation}. Finally, in the signal processing and performance evaluation module, the received signals are processed with different algorithms in order to estimate the parameters of the targets and analyze their relevant performance metrics. With the link-level simulation, the sensing accuracy performance can be calculated based on the estimated parameters and ground truth used for the simulation. An example of the link-level simulation with a ray tracing channel at 60 GHz in an indoor scenario is shown and discussed in \cite{11bf_lls_follow_up}, which verifies that WLAN sensing simulation can be conducted based on this link-level simulation platform.

\section{Conclusion}
	In this paper, we provided a comprehensive overview of the up-to-date efforts for the emerging IEEE 802.11bf standardization, including the new use cases, WLAN sensing procedure, candidate technical features, and evaluation methodology. Specifically, we initially introduced how IEEE 802.11bf was formed and the timeline for its standardization. Next, a detailed literature survey on use cases of WLAN sensing was provided. Furthermore, we described the sensing procedure used in IEEE 802.11bf to address the measurement acquisition problem for sensing. Then, several key candidate technical features that are likely to be approved in the IEEE 802.11bf were discussed and analyzed. Finally, we gave the evaluation methodology for the IEEE 802.11bf task group.

\end{document}